%% file: main.tex
\documentclass[10pt,twocolumn,letterpaper,table]{article}

\usepackage[pagenumbers]{iccv} %
\usepackage{xcolor}

\input{preamble}
\input{imports}

\input{definitions}

\definecolor{cvprblue}{rgb}{0.21,0.49,0.74}
\usepackage[pagebackref,breaklinks,colorlinks,allcolors=cvprblue]{hyperref}

\def\confName{ICCV}
\def\confYear{2025}

\title{\benchName{}: a Comprehensive Benchmark for Multi-view Generation Models}

\author{
Xianghui Xie$^{1,2,3}$\quad\quad
Chuhang Zou$^{4}$\quad\quad
Meher Gitika Karumuri$^{4}$ \quad\quad \\
Jan Eric Lenssen$^3$ \quad\quad
Gerard Pons-Moll$^{1,2,3}$ \\
\\
{\small $^1$University of T\"ubingen, Germany \hspace{1cm} $^2$T\"ubingen AI Center, Germany  } \\
{\small $^3$Max Planck Institute for Informatics, Saarland Informatic Campus, Germany\hspace{1cm} $^4$Independent researcher} \\
{\small\href{https://virtualhumans.mpi-inf.mpg.de/MVGBench/}{https://virtualhumans.mpi-inf.mpg.de/MVGBench/}}\\
}

\begin{document}
\twocolumn[{%
\renewcommand\twocolumn[1][]{#1}%

\maketitle
\begin{center}
    \centering
    \captionsetup{type=figure}
    \includegraphics[width=1.0\textwidth]{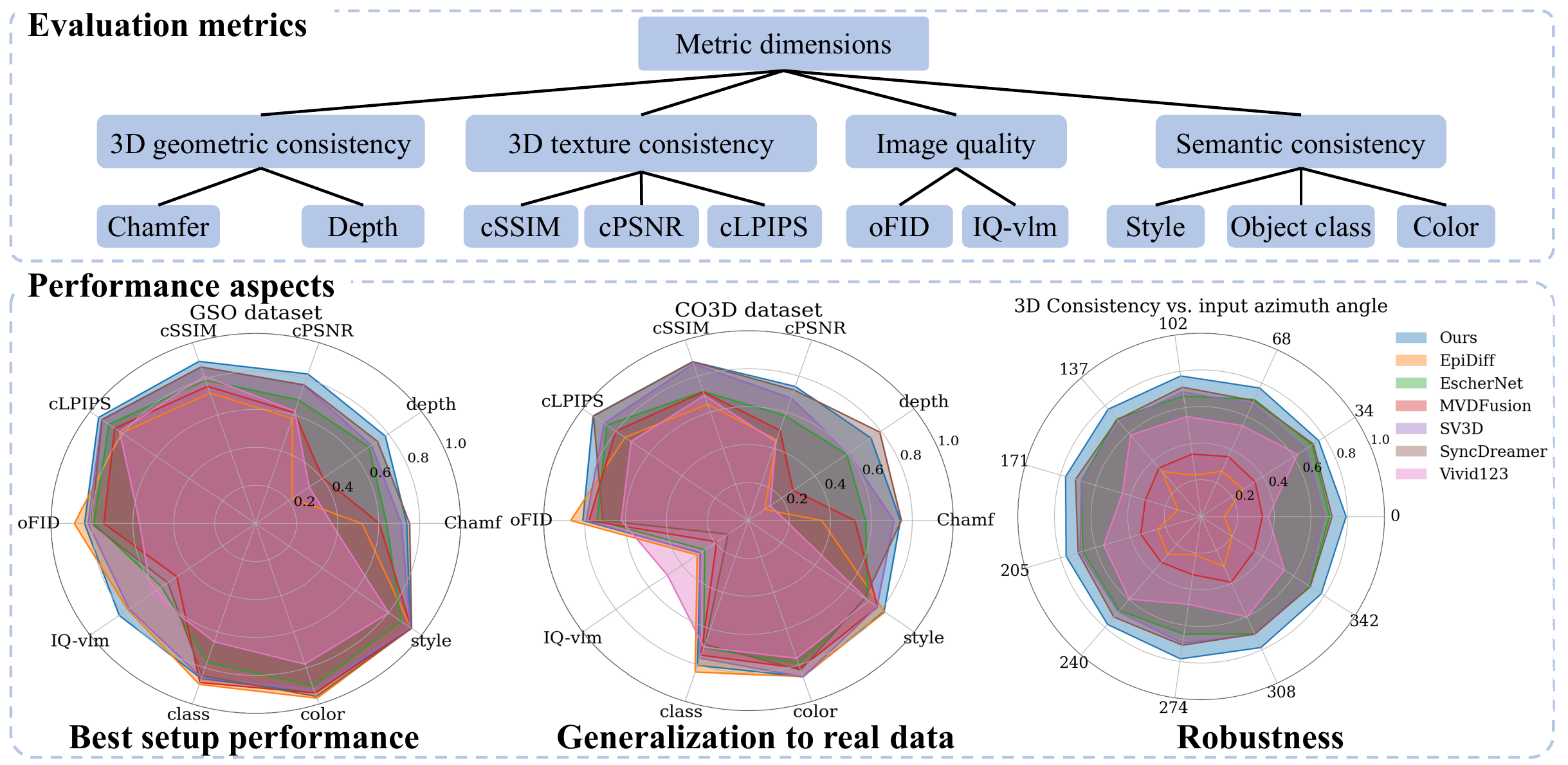}
    \captionof{figure}{
    We present \benchName{}, a comprehensive evaluation suite for multi-view image generation models (MVGs). We propose ten metrics to evaluate the 3D consistency in geometry and texture, image quality, and semantics of generated multi-view images. This suite allows us to fairly compare existing MVGs in three aspects: best setup performance, generalization, and robustness to input perturbations. We use our benchmark to systematically analyze different models and identify critical design choices, leading to a new model that achieves the best 3D consistency and robustness, with otherwise on-par performance. %
    All values are normalized, and outermost is better.
    }
    \label{fig:teaser}
\end{center}%
}]

\input{sec/0_abstract}    
\input{sec/1_intro}

\input{sec/2_related}

\input{sec/3_evaluation}

\input{sec/5_experiments}

\input{sec/6_conclusion}

\newpage
{
    \small
    \bibliographystyle{ieeenat_fullname}
    \bibliography{main}
}

\input{sec/X_suppl}

\end{document}

%% file: imports.tex
\usepackage{epsfig}
\usepackage{graphicx}
\usepackage{comment}
\usepackage{amsmath,amssymb} %
\usepackage{bbm}

\usepackage[font=small,skip=5pt]{caption}
\usepackage{float}

\usepackage{multirow}
\usepackage{tabularx}
\usepackage{soul}
\usepackage{array}
\usepackage{comment}
\usepackage{outlines}

\usepackage[mathscr]{euscript}
\newcolumntype{L}[1]{>{\raggedright\let\newline\\arraybackslash\hspace{0pt}}m{#1}}
\newcolumntype{C}[1]{>{\centering\let\newline\\arraybackslash\hspace{0pt}}m{#1}}
\newcolumntype{R}[1]{>{\raggedleft\let\newline\\arraybackslash\hspace{0pt}}m{#1}}
\newcolumntype{Y}{>{\centering\arraybackslash}X}
\usepackage{booktabs}
\usepackage{enumitem}

%% file: definitions.tex
\usepackage{amsmath}

\newcommand{\benchName}{MVGBench} %
\newcommand{\modelName}{ViFiGen} %
\newcommand{\qualityMetric}{IQ-vlm} %

\newcommand{\mat}[1]{\mathbf{#1}}
\newcommand{\set}[1]{\mathcal{#1}}

\definecolor{ForestGreen}{RGB}{14,109,14}

%% file: sec/0_abstract.tex
\begin{abstract}

\vspace{-0.2cm}

We propose \benchName{}, a comprehensive benchmark for multi-view image generation models (MVGs) that evaluates 3D consistency in geometry and texture, image quality, and semantics (using vision language models).
Recently, MVGs have been the main driving force in 3D object creation. However, existing metrics compare generated images against ground truth target views, which is not suitable for generative tasks where multiple solutions exist while differing from ground truth. 
Furthermore, different MVGs are trained on different view angles, synthetic data and specific lightings -- robustness to these factors and generalization to real data are rarely evaluated thoroughly. Without a rigorous evaluation protocol, it is also unclear what design choices contribute to the progress of MVGs. 

\benchName{} evaluates three different aspects: best setup performance, generalization to real data and robustness. Instead of comparing against ground truth, we introduce a novel 3D self-consistency metric which compares 3D reconstructions from disjoint generated multi-views. 
We systematically compare 12 existing MVGs on 4 different curated real and synthetic datasets. 
With our analysis, we identify important limitations of existing methods specially in terms of robustness and generalization, and we find the most critical design choices. 
Using the discovered best practices, we propose \modelName{}, a method that outperforms all evaluated MVGs on 3D consistency. Our benchmark suite and pretrained models will be publicly released.

\end{abstract}

%% file: sec/1_intro.tex
\section{Introduction}\label{sec:intro}

Powerful image generation models~\cite{rombach2021latentdiffusion, rombach2024stablediffusion_3_5, chang2023muse_t2i, VAR2024_t2i, llamagen2024_t2i} have been foundation models for a variety of tasks such as low-level vision~\cite{yang2023pasd_low_level_vision, wu2024osediff_low_level_vision, wang2023stablesr_low_level_vision}, image editing~\cite{zhang2023controlnet, zhange2023diffcollage, shi2023dragdiffusion, stableedit2023} and 3D generation~\cite{poole2022dreamfusion, wang2023prolificdreamer, liao2023tada, huang2024tech}. 
Multi-view generation (MVG) models~\cite{kongEscherNetGenerativeModel, voletiSV3DNovelMultiview2024a, liu2023syncdreamer, wang2023imagedream} which are trained to generate images at target views are particularly important as they have been the driving force for the rapid development of 3D content creation. Based on MVGs \cite{liu2023zero1to3, voletiSV3DNovelMultiview2024a, wang2023imagedream}, recent methods are able to create high quality 3D models from single images~\cite{xue2024gen3diffusion, xue2024human3diffusion, shi2023zero123plus, liu2023one2345} or text~\cite{shi2023MVDream, kantSPADSpatiallyAware, lu2024direct2_5}.

\begin{figure}
    \centering
    \includegraphics[width=1.0\linewidth]{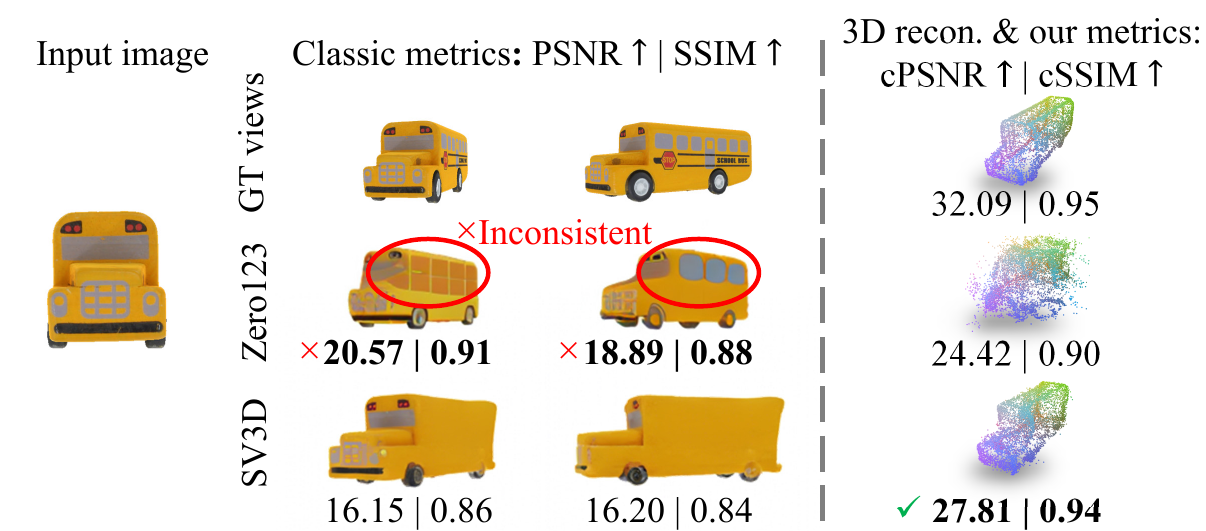}
    \caption{\textbf{Comparison of classic pair-wise metrics and our metrics}. Classic metrics compare generated images independently to paired ground truth views, which represent only one of many correct solutions in the ambiguous single-view generation task. In the example shown, despite the inconsistent generated multi-views, they assign a higher score to Zero123~\cite{stablezero123} while our metrics correctly identify SV3D~\cite{voletiSV3DNovelMultiview2024a} as the more 3D consistent method.}

    \label{fig:classic-metrics}
\end{figure}

Given the fundamental importance and rapid development of MVGs, proper evaluation is however lagging behind. Prior works~\cite{voletiSV3DNovelMultiview2024a, liu2023syncdreamer, kongEscherNetGenerativeModel} compute 2D metrics such as PSNR and SSIM between the generated and ground truth novel view images (\cref{fig:classic-metrics}). This is problematic for two reasons: a) the generative model samples from a distribution of solutions, so there is no single correct ground truth view; and b) the generated images are evaluated independently without considering 3D consistency. 
Furthermore, each method is trained using images with different rendering setups, resulting in different optimal input and output settings. Simply comparing them to method-specific ground truth will yield incomparable numbers, as the size of objects in 2D renderings varies between methods. 

Some works~\cite{wu2024direct3dgeneration, xue2024gen3diffusion, xue2024human3diffusion} compare the performance of MVGs by first lifting the multi-views to 3D, then aligning and calculating scores against 3D GT. The reliance on 3D GT makes it suboptimal to evaluate generative models and it is impossible to report numbers on real images where 3D GT is rarely available. Most works demonstrate selected examples, and only a few works conduct more rigorous evaluation via user studies, which is difficult to scale. 

To this end, we propose \benchName{}, a benchmark suite with comprehensive metrics and unified datasets for evaluating three important aspects of MVGs: \emph{a) Best setup performance.} The actual performance of each method using input images from their optimal camera setup. \emph{b) Generalization to real images.} We manually annotate real images with front view and elevation angles for a unified evaluation of generalization capabilities. \emph{c) Robustness to input perturbations.} We render objects at different elevations, azimuths and lighting conditions as input images and evaluate the performance of each method under these settings. 

A key contribution of our benchmark suite is a 3D consistency metric which does not compare against the ground truth and can faithfully assess methods that operate on different optimal setups. We evaluate 3D consistency by measuring the discrepancy between two 3D reconstructions obtained from two different subsets of the generated multi-view images. 
This design makes it suitable for evaluating generative models and also allows us to report quantitative results on real images in a more scalable way than user studies. 
In addition to 3D consistency, image quality and semantic consistency (class, color, and style) are also important for downstream applications, and we introduce metrics based on vision language models to evaluate these aspects.

We use our \benchName{} to evaluate 12 state-of-the-art MVGs and analyse the key design choices of the best performing methods. We observe that there is a trade-off between 3D consistency and image quality in existing models, and video diffusion models generally achieve a better balance. However, a significant performance gap persists between synthetic and real images, and most methods lack robustness to input perturbations and struggle with fine-grained details.
We further investigate design choices for 3D consistency and find that better camera embeddings and input image encoder can further improve performance. Leveraging best practices, we introduce \modelName{}, a video based multi-view generation method with a fine-grained input image encoder that outperforms all existing methods. Our evaluation suite and model will be publicly released.  

In summary, our \textbf{contributions} are:
\begin{itemize}
    \item We introduce the \benchName{} suite, with comprehensive metrics and manually curated datasets to evaluate MVGs. %
    \item We propose a novel 3D consistency metric that can fairly evaluate different MVGs on both synthetic and real data.
    \item  We use \benchName{} to systematically analyze and identify key problems of 12 state-of-the-art MVGs.
    \item We investigate the key design choices of the best performing MVGs and introduce \modelName{} that leverages the best practices and outperforms all existing baselines. 
\end{itemize}

%% file: sec/2_related.tex
\section{Related works}

\noindent\textbf{Multi-view image generation models (MVGs)} \cite{shi2023MVDream, wang2023imagedream, liu2023zero1to3, li2024era3d}, typically fine-tuned from large-scale image~\cite{rombach2021latentdiffusion, rombach2024stablediffusion_3_5} or video~\cite{blattmann2023videoldm, xing2023dynamicrafter} generation models, have exhibited strong generalization capability~\cite{hu2024turbo3d, xue2024gen3diffusion, xue2024human3diffusion} and significantly advanced 3D content creation~\cite{shi2023zero123plus, tang2024lgm, liu2023one2345, lu2024direct2_5, kantSPADSpatiallyAware, xu2024instantmesh}.   
Zero123~\cite{stablezero123} pioneered the repurposing of text to image generation model for camera-conditioned novel view synthesis. Follow-up works~\cite{shi2023MVDream, wang2023imagedream, shi2023zero123plus, zhengFree3DConsistentNovel2024, long2023wonder3d} further improve 3D consistency by simultaneously generating multiple views with advanced multi-view feature interaction mechanisms. 
SyncDreamer~\cite{liu2023syncdreamer} proposes to synchronize multi-view features via 3D convolutions, while video-based models~\cite{yang2024hi3d, voletiSV3DNovelMultiview2024a, li2024nvcomposer, chen2024v3d} rely on dense spatial-temporal attention driven by video data. 
Epipolar attention is also a common way to enhance the consistency between novel views~\cite{huangEpiDiffEnhancingMultiView, huMVDFusionSingleview3D, xuCamCo2024a}. Beside feature interaction, different camera embeddings~\cite{xuCamCo2024a, kongEscherNetGenerativeModel, zhengFree3DConsistentNovel2024, gao2024cat3d} and input image encoding~\cite{radford2021clip, Woo2023ConvNeXtV2, oquab2024dinov2} are also adopted. Despite promising results, there is no unified evaluation protocol for MVGs, making it difficult to understand the actual progress in this field and contributions of different design choices. Our benchmark suite allows for a unified evaluation and analysis of MVGs.

\noindent\textbf{Evaluation benchmarks or analysis} such as~\cite{cvpr10secrets_opticflow, benenson2014yearspedestriandetection, hodan2018bopbenchmark6dobject, bhatnagar22behave, xie2024rhobin} are essential to understanding the progress of a research field. The evaluation of generative models is less direct than traditional ground truth based evaluation and significant efforts have been made for evaluating image \cite{gao2024mvreward, xu2023imagereward, ku2024imagenhub}, video~\cite{liao2024evaluation, huang2024vbench, liu2023stylerf, asim2025met3r}, or 3D generation~\cite{wu2023gpteval3d}. These benchmarks focus on semantic consistency~\cite{ku2024imagenhub}, image or video quality~\cite{huang2024vbench}, video dynamics~\cite{liao2024evaluation} or alignment with human perceptions~\cite{gao2024mvreward, xu2023imagereward, wu2023gpteval3d}.    
However, in the field of MVG, most works~\cite{stablezero123, voletiSV3DNovelMultiview2024a, huMVDFusionSingleview3D, liu2023syncdreamer, kongEscherNetGenerativeModel, kwakViVid1to3NovelView, huang2024mvadapter, chen2024v3d, jeongNVSAdapterPlugandPlayNovel2024} still compare generated images against ground truth which is not meaningful and misses the important 3D consistency aspect, leading to inconsistent method rankings in different papers~\cite{voletiSV3DNovelMultiview2024a, liu2023syncdreamer, huang2024mvadapter}. Flow Warping Score~\cite{liu2023stylerf} and concurrent work MEt3R~\cite{asim2025met3r} measure consistency but focus on 3D scene generation. Some consistency metrics~\cite{zhengFree3DConsistentNovel2024, watsonNovelViewSynthesis2022} are proposed for object-level MVGs but are not well adopted. There is no unified and reliable evaluation protocol for object-level MVG models, let alone comprehensive analysis of this fast-evolving field with more than 20 papers per year. Our \benchName{} provides a reliable 3D consistency metric and the first unified benchmark framework that allows fair comparison and systematic analysis.

%% file: sec/3_evaluation.tex
\section{\benchName{} Evaluation Suite}
\definecolor{myPink}{HTML}{E91E63}  %
\definecolor{myBlue}{HTML}{2196F3}  %
\definecolor{myGreen}{HTML}{4CAF50} %
\begin{figure}[t]
    \centering
    \includegraphics[width=1.0\linewidth]{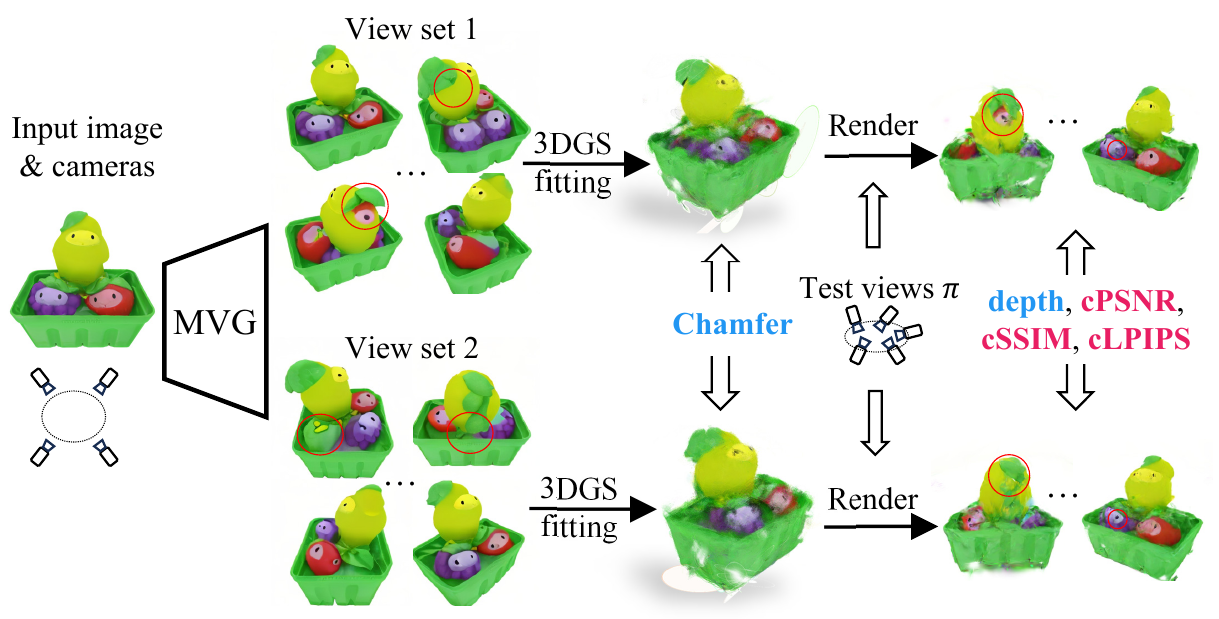}
    \caption{\textbf{3D consistency metrics for multi-view generation models}. 
    After prompting the model to generate multi-views at target camera poses, we split the output views and fit 3D Gaussian Splatting (3DGS) separately into two view sets. We measure the \textbf{\textcolor{myBlue}{geometric}} and \textbf{\textcolor{myPink}{texture}} consistency between two 3DGSs as the 3D consistency of the multi-view generation model.
    }
    \label{fig:consistency-pipeline}
\end{figure}

We present \benchName{}, a comprehensive evaluation suite for benchmarking multi-view generation models. We focus on evaluating models that generate images of a single or compositional object in the center while the background is masked out. Our evaluation suite consists of comprehensive evaluation metrics, including 3D consistency (\cref{subsec:3d-consistency-metric}), image quality, and semantic consistency (\cref{subsec:semantic-quality}). We then curate several datasets to evaluate methods on three distinct aspects (\cref{subsec:eval-datasets}). An overview of our metric dimensions and performance aspects can be found in \cref{fig:teaser}. Please see metric implementation details in our Supp.

\subsection{3D Consistency Metrics}\label{subsec:3d-consistency-metric}
Generated multi-view images should be 3D consistent to form a coherent 3D model~\cite{xue2024human3diffusion, xue2024gen3diffusion}. Previous methods evaluate this by first reconstructing 3D from multi-view images and then comparing to the 3D ground truth. However, this cannot be scaled to datasets without 3D 
ground truth. %
Moreover, generative models can produce images that differ from ground truth but are consistent, making this metric unreliable.
Our key idea is to measure 3D consistency via self-consistency in 3D between generated images. An overview of our 3D consistency metrics is shown in \cref{fig:consistency-pipeline}. 

Given a single RGB image $\mat{I}$ and $N$ target view camera poses, we prompt the multi-view generation model to generate $N$ images $\set{I}=\{\mat{I}_1,... \mat{I}_N\}$ at target views. We then divide these images into two subsets $\set{I}_1=\{\mat{I}_1^1, ... \mat{I}_n^1\}, \set{I}_2=\{\mat{I}_1^2, ... \mat{I}_n^2\}$ and fit two 3D Gaussian Splattings (3DGSs~\cite{kerbl3Dgaussians}) to them separately. We allow a small view overlap between $\set{I}_1$ and $\set{I}_2$ when $N$ is small (detailed later). 
Let $\set{G}_1, \set{G}_2$ denote the optimized 3DGSs from two view sets $\set{I}_1, \set{I}_2$ respectively. In principle, the two 3DGSs will be very similar if the generated images are consistent with each other. Therefore, we measure the discrepancy between two 3DGSs as the indicator for the 3D geometric and texture consistency of the generated multi-views. %

\noindent\textbf{Geometric consistency metrics} evaluate the consistency of the geometric structure between two 3DGSs fitted separately from two subsets of multi-views.
We compute the \texttt{Chamfer} distance (CD) and rendered \texttt{depth} error between two 3DGSs. The Chamfer distance $e_\text{cd}$ measures the discrepancy in the overall shape structure, while the depth error $e_d$ is more sensitive to edge inconsistencies. 

For the \textbf{Chamfer distance}, instead of using the Gaussian centers that do not faithfully represent the actual surface of the shape, we resample the 3DGSs using the optimized covariance matrices. 
Let $\mat{\mu}_1\in\mathbb{R}^{M_1\times 3}, \mat{\Sigma}_1\in\mathbb{R}^{M_1\times 3\times 3}$ be the centers and covariances of the 3DGS $\set{G}_1$ from images $\set{I}_1$, the Chamfer distance between two 3DGSs is defined as:
\begin{align}
    e_\text{cd}(\set{G}_1, \set{G}_2) &= d_\text{CD}(\mat{P}_1, \mat{P}_2) \\
    \text{ where }\mat{P}_1&\sim \set{N}(\mat{\mu}_1, \mat{\Sigma}_1), \mat{P}_2\sim \set{N}(\mat{\mu}_2, \mat{\Sigma}_2)
    \label{eq:metric-chamfer}
\end{align}
here $d_\text{CD}(\cdot, \cdot)$ is the Chamfer distance between two sets of points. Two 3DGSs usually have different number of Gaussians. Hence we first sample five points from each Gaussian as $\mat{P}_1, \mat{P}_2$ and then downsample  to 60k points to compute $d_\text{CD}$. 
We empirically found that this resampling produces more faithful distance values. 

For the \textbf{depth error}, we render $K$ depth maps of optimized 3DGSs following LGM~\cite{tang2024lgm}, using the same camera views for two 3DGSs. We then compute the error between the two depth maps after masking out empty background: 
\begin{equation}
    e_d(\mathcal{G}_1, \mathcal{G}_2)=\frac{1}{K}\sum_{i=1}^K \frac{1}{|\mat{M}_i|}\sum \mat{M}_i|\pi^d_i(\mathcal{G}_1)-\pi^d_i(\mathcal{G}_2)|
    \label{eq:metric-depth}
\end{equation}
where $\pi^d_i$ denotes the depth rendering of view $i$ and $\mat{M}=(\pi^d_i(\mathcal{G}_1)>0)|(\pi^d_i(\mathcal{G}_2)>0)$ is the union mask of the two rendered foregrounds. 

\textbf{Texture consistency metrics.} %
We render each of the two optimized 3DGSs $\set{G}_1, \set{G}_2$ into $K$ different views using same cameras as $\pi^d_i$. We then calculate the distance between the two renderings using PSNR, SSIM, and LPIPS:
\begin{equation}
    e_m(\set{G}_1, \set{G}_2) = \frac{1}{K}\sum_{i=1}^K d_m (\pi_i(\set{G}_1), \pi_i(\set{G}_2))
    \label{eq:metric-texture}
\end{equation}
where $\pi_i$ renders 3DGS $\set{G}$ into an RGB image and $d_m(\cdot, \cdot)$ is the distance between two images defined by PSNR, SSIM or LPIPS. %
To distinguish them from classic names, we call these consistency metrics \texttt{cPSNR}, \texttt{cSSIM}, and \texttt{cLPIPS}. 

\textbf{Handling different input and output setups.} One challenge in comparing different MVGs is that these methods render the training images at different camera focal lengths, distances, or elevations. This creates training data bias, and each model works optimally in different setups. 
Therefore, it would be \emph{unfair} to render the same input image for all methods and compare the results, as this would favor methods trained on a similar camera setup. Using different testing inputs yields output objects of different sizes or view angles, and comparing against method-specific GT views (as done in prior evaluation) leads to incomparable numbers across methods. 
Our consistency metrics address this problem by design, as we can define arbitrary rendering views $\pi_i$ for evaluation and use the same view across all methods. The only requirement is that the optimized 3DGSs are aligned in 3D space, as we discuss below. 

For synthetic datasets where 3D is available, we normalize the 3D object within the unit cube and use the training camera setup (focal, distance) of each method to render its input image. When prompted with this image, the generated multi-views should represent a 3D object that aligns with the normalized 3D object in scale and rotation. Hence, the 3DGSs optimized with each methods' own cameras are also aligned, allowing us to define same test views $\pi_i$ and compute scores that are comparable. 
For real images where we cannot recreate the input image at a specific camera setup, the optimized 3DGS will be misaligned since different methods generate different object sizes given the same input image. In this case, we use ICP with uniform scaling to align 3DGSs from different methods to a reference 3DGS from one MVG. We then use the same camera views for all methods to render images from the aligned 3DGS, leading to aligned consistency scores, as desired. 

The number of output views each method is trained on also differ, affecting the accuracy of 3DGS fitting. To align the errors raised from 3DGS fitting given different views, we allow small overlap between the two view sets when the number of generated views is small, leading to a fair upper bound given different number of GT multi-views (\cref{tab:val-consistency-metric}). %

\textbf{Discussions.} Our 3D consistency metric has two advantages over traditional evaluation metrics: (1) No comparison with 3D GT, which makes it more suitable for evaluating generative models and reporting quantitative results on real images. (2) Fair comparison of methods. Our metric allows each method to take input rendered in its own training data setting and report numbers that still align.

\begin{figure*}[t]
    \centering
    \includegraphics[width=1.0\linewidth]{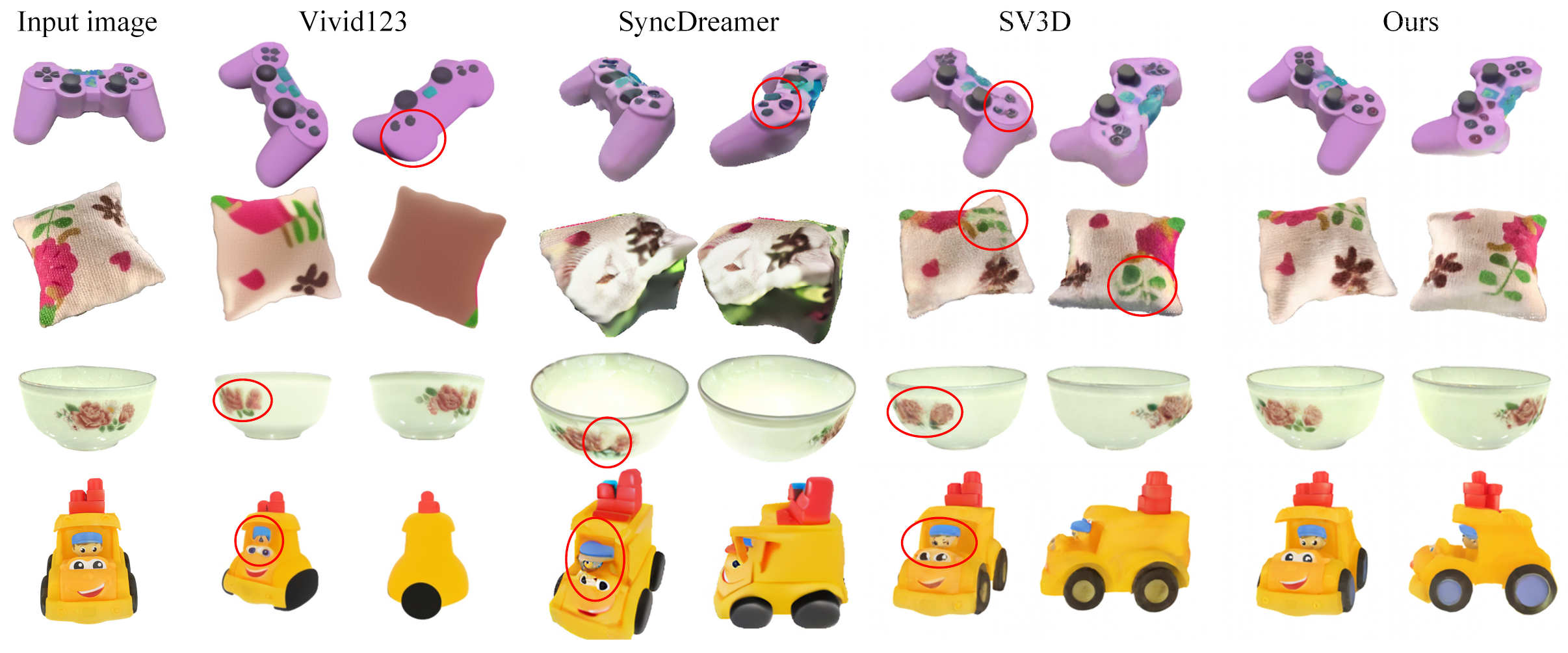}
    \vspace{-1.9em}
    \caption{\textbf{Example results} on CO3D~\cite{reizenstein2021co3d} (row 1), MVImgnet~\cite{yu2023mvimgnet} (row 2), Omni3D~\cite{wu2023omniobject3d} (row 3) and GSO~\cite{downs2022gso} (row 4). Vivid123~\cite{kwakViVid1to3NovelView} generates images that look good but are 3D inconsistent while SyncDreamer~\cite{liu2023syncdreamer} images are 3D consistent but the image quality is worse. SV3D~\cite{voletiSV3DNovelMultiview2024a} and our model adopts video prior and achieves a better trade-off between consistency and quality. Our method leverages ConvNextV2~\cite{Woo2023ConvNeXtV2} instead of CLIP~\cite{radford2021clip} to encode input image and preserves better details than SV3D.}
    \label{fig:qualitative}
    \vspace{-1.3em}
\end{figure*}

\subsection{Semantic and Image Quality Metrics}\label{subsec:semantic-quality}
The 3D consistency metric in \cref{subsec:3d-consistency-metric} is suitable for evaluating the potential of generated images for 3D tasks. However, a list of pure white images is perfectly 3D-consistent but semantically meaningless. Therefore, it is also valuable to measure semantics which are orthogonal to 3D consistency but still important. To this end, we propose five metrics to measure image quality and semantic consistency.

\textbf{Image quality.} We assess the quality of generated multi-views using average object FID (\texttt{oFID}) and pretrained vision language models (VLMs). 
We compute oFID using CLIP features~\cite{radford2021clip} which is more robust to diverse objects~\cite{Jayasumana2024RethinkingFID}. Let $\mathcal{F}^\text{gt}_i, \mathcal{F}_i$ be the CLIP features of the object $i$ from $N$ target views, oFID is the average FID of all objects in the test dataset $\mathcal{D}$: $\text{oFID}=\sum_{i=1}^{|\set{D}|}\text{FID}(\set{F}_i^\text{gt}, \set{F}_i)$, where $\text{FID}(\cdot, \cdot)$ is the standard Fr\'echet distance. We show in \cref{subsec:validate-metrics} that our oFID aligns better with human preferences than the classic dataset level FID. Rendering at target views is difficult for real data as 3D GT is typically not available, we hence use datasets that capture each object in a video and randomly sample frames from this video to extract GT multi-view features.

We further prompt a pretrained VLM~\cite{chen2024internvl_2_5} to assess the overall image quality of the generated images and return a binary ``yes" or ``no" answer (yes means overall good). The score is the percentage of ``yes" returned by the VLM, denoted as \texttt{\qualityMetric{}} (image quality via VLM).  

\textbf{Semantic consistency.} We use the pretrained VLM~\cite{chen2024internvl_2_5} to assess whether the generated images are semantically consistent with the input image. We first prompt the VLM to annotate semantic attributes (object class, color, and appearance style) of each input using ground truth multi-view images, yielding the reference attributes. We then prompt the VLM with the generated images and ask if the reference attribute is presented in the image. Each attribute evaluation is a binary ``yes/no" question with ``yes" meaning the attribute is presented in the image. We denote these three metrics as \texttt{class}, \texttt{color}, and \texttt{style}. 
We show in \cref{subsec:validate-metrics} that our VLM-based metrics align well with human preferences and prompt template in Supp. 

\subsection{Evaluation Datasets}\label{subsec:eval-datasets}
To have a common benchmark for comparing different methods, we curated several datasets and further processed them to evaluate three important performance aspects: best setup performance, generalization, and robustness. %

\textbf{Best setup performance.} This setup is designed to compare the best performance of each method. Inputs are rendered from 3D models using each method's own training setup, ensuring that each method is run under optimal input conditions. %
We use Google Scanned Objects (GSO\cite{downs2022gso}) and OmniObject3D (Omni3D\cite{wu2023omniobject3d}) for this aspect. 
For GSO, we reuse 30 objects used in previous works (GSO30~\cite{liu2023syncdreamer, kongEscherNetGenerativeModel, voletiSV3DNovelMultiview2024a}) and sample 70 more non-overlapping objects. For Omni3D, we randomly sample one instance from each category, resulting in 202 unique objects.

\textbf{Generalization to real images.} %
Existing methods typically show different qualitative examples of selected images, making it difficult to compare actual generalization abilities. Conducting user studies is more rigorous, but does not scale. Our metrics allow us to quantify the generalization performance on real images, which is more scalable. 
We use MVImgnet~\cite{yu2023mvimgnet} and CO3D~\cite{reizenstein2021co3d} for this evaluation. Both datasets capture multi-view images of real objects with unaligned camera poses, while most methods~\cite{voletiSV3DNovelMultiview2024a, liu2023syncdreamer, huMVDFusionSingleview3D} require elevation angle as input. To this end, we manually select images that best fit as the frontal view and annotate their elevation angle. 
We select two instances per category for CO3D and one instance per category for MVImgnet, leading to 102 and 230 images, respectively. 

\textbf{Robustness.} Another important aspect is to understand the robustness of the model under different input perturbations, which is barely done in previous evaluations. We consider three types of input perturbations that are difficult to undo via input image processing once the images are captured: elevation, azimuth, and light intensity. 
To do this, we render the GSO30 objects used by ~\cite{kongEscherNetGenerativeModel} under different conditions as input images. Note that in addition to the control factors we consider, we still use the camera focal length and distance from the training setup of each method to ensure that each method operates under optimal condition.

Most MVGs generate images at the same elevation as input, hence the optimized 3DGS is mostly accurate at similar elevations. Errors in renderings far from this elevation are mainly caused by 3DGS fit rather than multi-view inconsistencies. This is problematic when evaluating robustness w.r.t. different input elevations, since it is not possible to have the same test views for all input elevations. We hence use test views with a 15 degree offset the from input elevation, and normalize the consistency scores using an upper bound defined by the ground truth multi-view images. See the appendix for the normalization formula.

%% file: sec/5_experiments.tex
\section{Experiments}
In this section, we first validate our proposed metrics and then present our evaluation analysis of 12 typical MVGs. We then further systematically investigate some key design choices of MVGs and propose a method that outperforms all existing baselines in terms of 3D consistency.

\subsection{Validating \benchName{} metrics}\label{subsec:validate-metrics}
\textbf{3D Consistency.} 
A good 3D consistency metric should be invariant to the number of views used and the specific camera setups. 
Therefore, using ground truth views (perfectly 3D consistent), we report our 3D consistency metrics varying these aspects.
As can be seen from \cref{tab:val-consistency-metric}, the deviation is less than 8\% of the average score across all variants of our metric. Note that the score differs from the theoretical upper bound due to slight inaccuracies in 3DGS fitting. 
However, these numbers indicate that such inaccuracies are negligible and are not affected by specific method setups. 
Hence, we can conclude that our metric scores vary only due to the 3D inconsistency of multi-view images. 
\begin{table}[h]
    \centering
    \vspace{-0.5em}
    \footnotesize
    \begin{tabular}{c c |>{\centering\arraybackslash}p{0.65cm} >{\centering\arraybackslash}p{0.65cm} |>{\centering\arraybackslash}p{0.65cm} >{\centering\arraybackslash}p{0.65cm} >{\centering\arraybackslash}p{0.7cm} }
         \#views & camera & CD$\downarrow$ & depth$\downarrow$ & cPSNR$\uparrow$& cSSIM$\uparrow$& cLPIPS$\downarrow$ \\
         \hline
         16 & \cite{huMVDFusionSingleview3D}  & 1.993 & 6.941 & 30.281 & 0.924 & 0.034 \\ 
         18 & \cite{chen2024v3d} & 2.119 & 7.974 & 30.688 & 0.934 & 0.030  \\ 
         20 & \cite{liu2023zero1to3} & 2.133 & 6.977 & 30.448 & 0.925 & 0.031 \\ 
         20 & \cite{voletiSV3DNovelMultiview2024a} & 2.091 & 6.350 & 30.800 & 0.932 & 0.029  \\ 
         \hline 
         \multicolumn{2}{c|}{relative std.} & 0.026 & 0.082 & 0.006 & 0.004 & 0.051 \\ 
    \end{tabular}
    \caption{Our 3D consistency metrics are \textbf{invariant to number of views and different camera rendering settings}. %
    }
    \vspace{-0.5em}
    \label{tab:val-consistency-metric}
\end{table}

\begin{figure}
    \centering
    \includegraphics[width=1.0\linewidth]{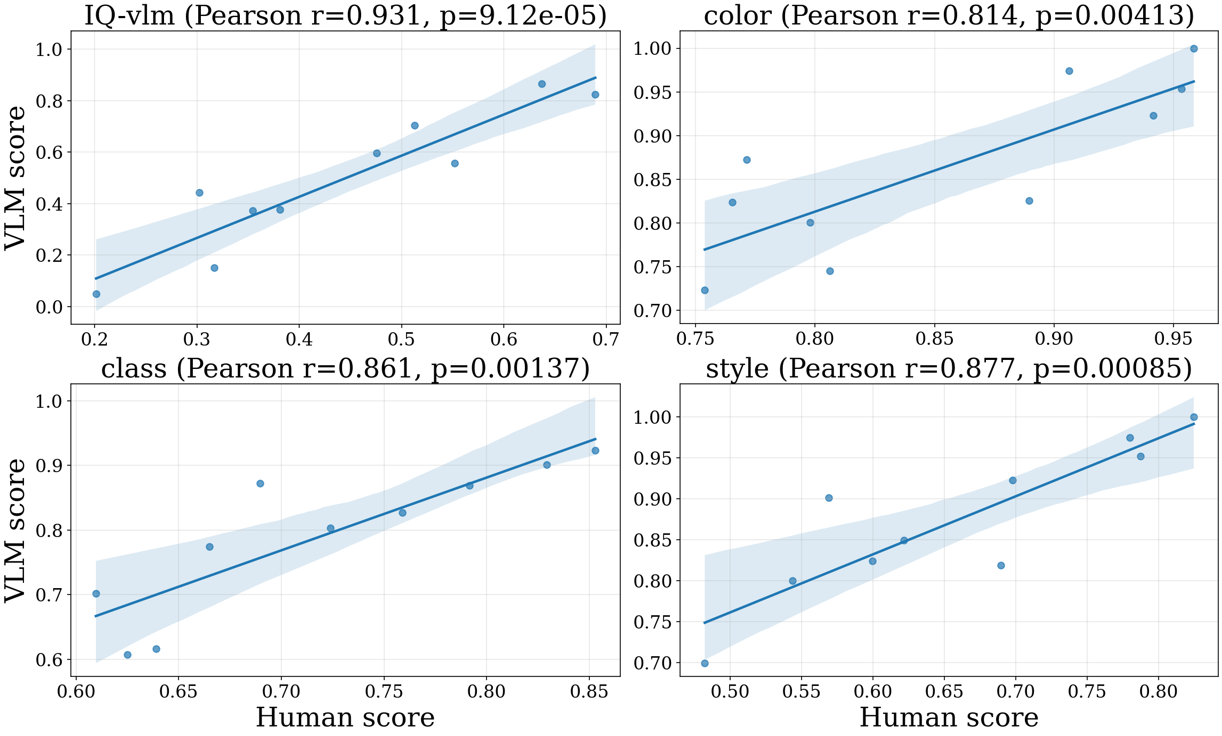}
    \caption{\textbf{Validating vision language model (VLM) based metrics.} Our VLM metrics strongly correlate with human perception (Pearson coefficient confidence interval: 0.95).}
    \label{fig:vlm-pearson}
\end{figure}

\noindent\textbf{VLM metrics.} We propose four metrics based on a pretrained VLM~\cite{chen2024internvl_2_5} to evaluate image quality and semantic consistency. We conduct a user study to verify if they align with human perception. We randomly select 400 generated images from 5 methods in GSO~\cite{downs2022gso} and CO3D~\cite{reizenstein2021co3d} datasets and ask 10 
users to answer the exact same verification questions as the VLM (yes means the image passes the check). 
Notably, the average scores (percentage of yes) reported by the users strongly correlate with the one from the VLM, see \cref{fig:vlm-pearson}. We can hence conclude that our proposed VLM metric is faithful to human perception. 

\noindent
\textbf{oFID score.} 
We propose oFID score that measures how well the generated image of a specific object matches the distribution of multi-views of that object. 
Hence, instead of computing FID against the full dataset of all objects, we compute FID per object instance, and then average.
To verify that our proposed oFID aligns better with human perception, we perform user studies to compare the method rankings based on oFID, FID and humans. 
We ask users to rank methods pair-wise based on alignment with the input image. 
We select 10 pairs of methods and for each pair, 40 random examples are selected and evaluated by 10 users. 
 We report the percentage of ranking matches between FID or our oFID and human rankings: 0.77 (oFID), 0.50 (dataset FID). Clearly, oFID is much better aligned with human rankings. Our oFID and human rankings are strongly correlated and have a Pearson and Spearman coefficient of 0.69 and 0.67. We show example user study questions in the appendix.

\subsection{Evaluation results and observations}\label{subsec:exp-eval}

\subsubsection{Evaluation setup}
We evaluate open-sourced multi-view generation methods for object-level image generation. We exclude methods that only generate fixed few-views such as ImageDream~\cite{wang2023imagedream}, Zero123++~\cite{shi2023zero123plus}, Wonder3D~\cite{long2023wonder3d} and \cite{wuDirectExplicit3D2024}. Hence, we evaluate the following 12 SoTA methods: Zero123~\cite{liu2023zero1to3}, zero123-xl~\cite{deitke2023objaversexl}, Vivid123~\cite{kwakViVid1to3NovelView}, EpiDiff~\cite{huangEpiDiffEnhancingMultiView}, Free3D~\cite{zhengFree3DConsistentNovel2024}, MVDFusion~\cite{huMVDFusionSingleview3D}, ViewFusion~\cite{yangViewFusionMultiViewConsistency}, EscherNet~\cite{kongEscherNetGenerativeModel}, SyncDreamer~\cite{liu2023syncdreamer}, V3D~\cite{chen2024v3d}, Hi3D~\cite{yang2024hi3d}, SV3D\cite{voletiSV3DNovelMultiview2024a}. See the appendix for details about the individual input and output setups. We also add a new baseline method that leverages the best practices of MVGs, named ours (detailed in \cref{subsec:exp-mvg-design}).

\subsubsection{Observations}
We present the summarized analysis in this section. The full evaluation tables can be found in the appendix. 

\begin{figure}[]
    \centering
    \vspace{-1em}
    \includegraphics[width=1.0\linewidth]{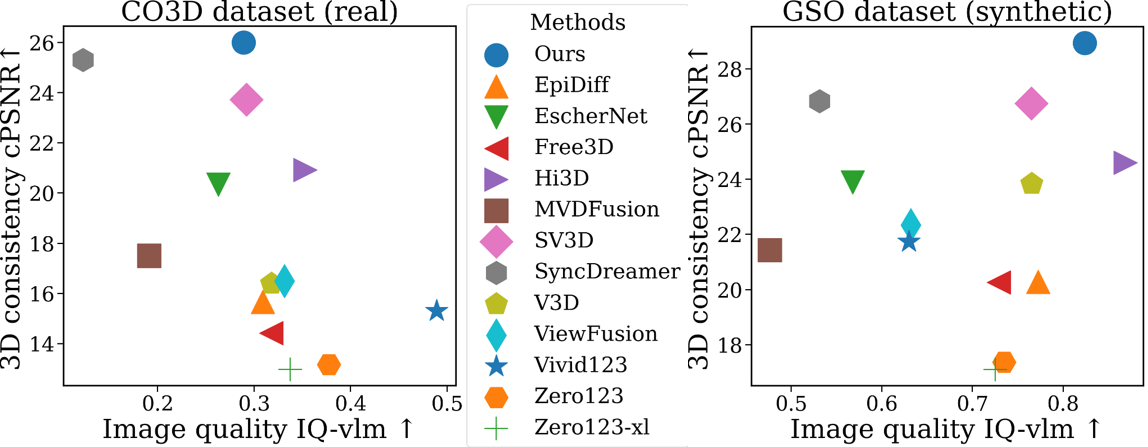}
    \caption{\textbf{Trade-off between 3D consistency and image quality}. No method can achieve the best performance in both dimensions.}
    \label{fig:trade-off}
\end{figure}

\noindent\textbf{Trade-off between 3D consistency and quality.} We plot the 3D consistency (cPSNR) versus image quality (IQ-vlm) on our CO3D~\cite{reizenstein2021co3d} and GSO~\cite{downs2022gso} test set in \cref{fig:trade-off}. It can be seen that the top-right (best in both consistency and quality) region is unoccupied. Methods like Zero123~\cite{liu2023zero1to3}, Vivid123~\cite{kwakViVid1to3NovelView} produce images that achieve good image quality, but are not 3D consistent. Conversely, methods like SyncDreamer~\cite{liu2023syncdreamer}, EscherNet~\cite{kongEscherNetGenerativeModel} and MVDFusion~\cite{huMVDFusionSingleview3D} can generate 3D consistent images at the cost of lower quality. 
We observe that methods are either consistent but lack detail, or have detail that improve quality but difficult to be consistent. 
More recent video-based methods SV3D~\cite{voletiSV3DNovelMultiview2024a} and ours balance 3D consistency and quality better, see example images in \cref{fig:qualitative}. It is also visible in \cref{fig:teaser} where no method can reach the best score on all dimensions. 

\noindent\textbf{Large gap between synthetic and real images.} We report the key metrics in 3D consistency (geometry+texture) and image quality (IQ-vlm) of all methods in \cref{tab:gso-vs-co3d}. The performance on real images (CO3D~\cite{reizenstein2021co3d}) is considerably inferior to the performance on synthetic images (GSO~\cite{downs2022gso}), especially in the  (IQ-vlm). We also show comparisons in \cref{fig:qualitative} where the generated images are much worse for real images. More examples are presented in the appendix. 

\begin{table}[]
    \centering
    \scriptsize
    \begin{tabular}{p{1.5cm}|>{\centering\arraybackslash}p{0.55cm} >{\centering\arraybackslash}p{0.70cm}>{\centering\arraybackslash}p{0.70cm}|>{\centering\arraybackslash}p{0.55cm}>{\centering\arraybackslash}p{0.70cm}>{\centering\arraybackslash}p{0.70cm}}
    \multirow{2}{*}{Method} & \multicolumn{3}{c}{CO3D~\cite{reizenstein2021co3d} (real)} & \multicolumn{3}{c}{GSO~\cite{downs2022gso} (synthetic)} \\
      & {\tiny CD $\downarrow$}& {\tiny cPSNR $\uparrow$} & {\tiny \qualityMetric{} $\uparrow$} & {\tiny CD $\downarrow$}& {\tiny cPSNR $\uparrow$} & {\tiny \qualityMetric{} $\uparrow$} \\
      \hline
         Ours & {\bf 3.02} & {\bf 25.82} & 0.29 & 3.15 & {\bf 28.93} & 0.82 \\
         SV3D-tune & 3.32 & 24.70 & 0.29 & 3.24 & 27.95 & 0.82  \\
         SyncDreamer & 3.04 & 25.30 & 0.12 & {\bf 2.99} & 26.83 & 0.53  \\
         SV3D & 3.48 & 23.72 & 0.29 & 3.47 & 26.75 & 0.77 \\
Hi3D & 5.60 & 20.92 & 0.35  & 3.29 & 24.60 & {\bf 0.87}  \\
Eschernet& 5.14 & 20.34 & 0.26  & 4.34 & 23.89 & 0.57 \\
V3D & 12.24 & 15.87 & 0.32& 4.25 & 23.83 & 0.81  \\
ViewFusion & 10.45 & 16.39 & 0.33& 5.33 & 22.34 & 0.63  \\
MVDFusion & 5.77 & 17.50 & 0.19 & 4.77 & 21.44 & 0.48  \\
Free3d & 11.15 & 14.42 & 0.32  & 6.03 & 20.26 & 0.73 \\
EpiDiff & 7.71 & 15.66  & 0.31 & 5.77 & 20.28 & 0.77  \\
Vivid123 & 9.81 & 15.31 & {\bf 0.49}& 7.57 & 21.74 & 0.63  \\
Zero123 & 12.06 & 13.16 & 0.38 & 10.99 & 17.37 & 0.73  \\
Zero123-xl & 12.58 & 12.97 & 0.34  & 15.40 & 17.10 & 0.72 \\
    \end{tabular}
    \caption{\textbf{Evaluation results on GSO and CO3D}. The performance gap between synthetic and real data is large, especially on the image quality aspect (IQ-vlm).}
    \label{tab:gso-vs-co3d}
\end{table}

\begin{figure*}[th]
    \centering
    \includegraphics[width=1.0\linewidth]{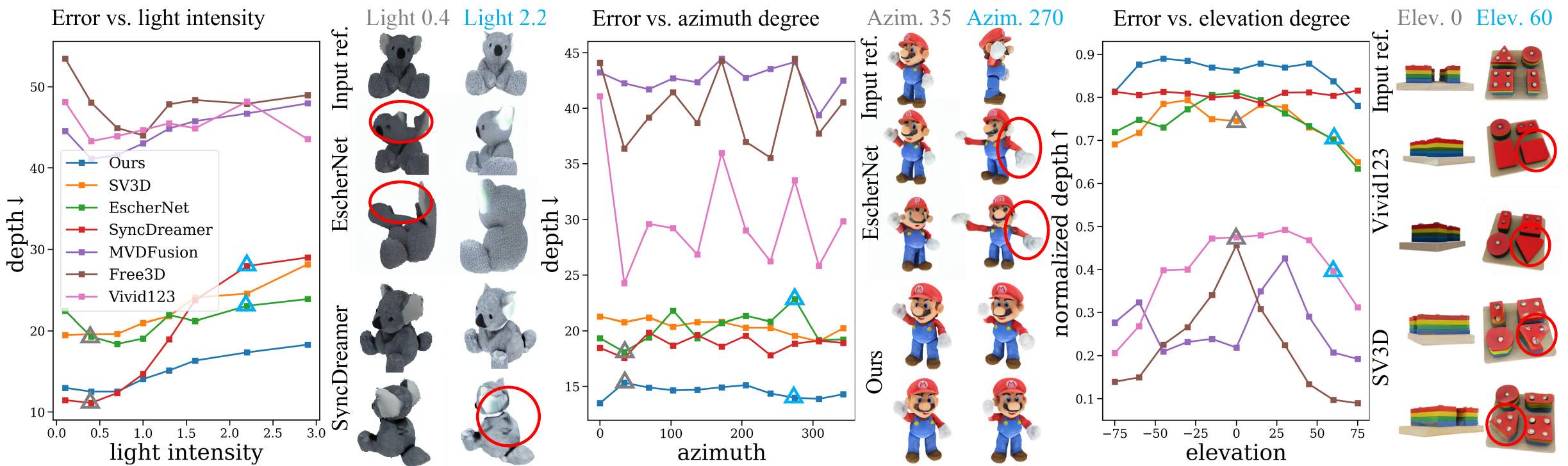}
    \caption{\textbf{Robustness w.r.t different light intensity, azimuth and elevation angles.} Some methods (EscherNet~\cite{kongEscherNetGenerativeModel}) are sensitive to dark lighting while others (SyncDreamer~\cite{liu2023syncdreamer}) are sensitive to strong lighting. Some methods (EscherNet, Vivid123~\cite{kwakViVid1to3NovelView}) are also sensitive to the input azimuth angles and none of the methods are robust to higher elevation angles.}
    \label{fig:robostness-all}
    \vspace{-1em}
\end{figure*}

\noindent\textbf{Methods are not robust to input perturbations.} We plot the error versus different input light intensity, azimuth and elevation degrees, alongside representative examples in \cref{fig:robostness-all}. %
Some methods are sensitive to dark light or specific azimuth and none of them are robust to elevations. We attribute this to limited pose variation in training. In contrast, our method trained on diverse renderings is more robust especially wrt. different lighting and azimuth angles.

\noindent\textbf{Methods struggle to handle fine structures or textures.} 
To identify the most challenging images for each method, we rank the input images based on 3D consistency score. Objects with complex and fine-grained geometric structures or textures such as bicycle, flowers, text boxes are the most challenging. One reason is that the autoencoder for latent diffusion~\cite{rombach2021latentdiffusion} destroys high-frequency details and simply passing images through the autoencoder already degrades 3D consistency, see analysis in the appendix.

\subsection{MVG design choices}\label{subsec:exp-mvg-design}
What makes a MVG 3D consistent? With our benchmark suite, we can study this question from a fair and unified perspective. 
We classify different design choices into four categories: \textbf{a) Camera pose embedding.} Most methods~\cite{liu2023zero1to3, voletiSV3DNovelMultiview2024a} adopt simple MLP based embedding while EscherNet proposes to use CaPE and Free3D adopts Plucker ray-based embedding.  \textbf{b) Input image encoder.} Most methods~\cite{liu2023zero1to3, liu2023syncdreamer, voletiSV3DNovelMultiview2024a} use CLIP~\cite{radford2021clip} to encode input image while EscherNet~\cite{kongEscherNetGenerativeModel} adopts ConvNextV2~\cite{Woo2023ConvNeXtV2} encoder to extract fine-grained features. We also consider the DINOv2~\cite{oquab2024dinov2} encoder. \textbf{c) Interaction between target view features.} Recent methods generate multiple target views together and compute different attentions between target view features. Due to resource limits, we only compare the synchronized 3D convolution from SyncDreamer~\cite{liu2023syncdreamer} and spatial-temporal attention from SV3D~\cite{voletiSV3DNovelMultiview2024a}. \textbf{d) How much training data is needed} to train a good MVG? The training data used in prior methods range from 40k to 800k objects but it is unclear how much is actually needed. 

We adopt SV3D as our base model, which balances well between consistency and quality, and study the contribution of these different design choices. SV3D uses simple MLP to embed camera pose and the input image is encoded with CLIP. It generates 21 images together and uses the spatial-temporal attention pretrained for video generation. 

\noindent\textbf{Camera embedding.} We add the camera positional encoding (CaP) from either EscherNet~\cite{kongEscherNetGenerativeModel} or the ray conditional network (RCN) from Free3D~\cite{zhengFree3DConsistentNovel2024} to SV3D. We fine tune the modified network for 26k steps. To rule out the effect of training data, we also fine tune SV3D in the same dataset and the results are reported in \cref{tab:ablate-architecture} b-e. It can be seen that both Cap (\cref{tab:ablate-architecture}d) and RCN 
(\cref{tab:ablate-architecture}c) are more effective than simple MLP based embedding (\cref{tab:ablate-architecture}b). RCN is better than CaP and interestingly, combing both leads to a worse result. However, when replacing CLIP with convolution encoder, the difference bewteen CaP (\cref{tab:ablate-architecture}h) and RCN (\cref{tab:ablate-architecture}f) is small. 

\definecolor{myTeal}{HTML}{1ABC9C}   %
\definecolor{myPurple}{HTML}{9B59B6} %
\definecolor{myOrange}{HTML}{F39C12} %
\colorlet{tealPurpleBlend}{myTeal!50!myPurple}
\colorlet{tealOrangeBlend}{myTeal!50!myOrange}

\definecolor{myPink}{HTML}{E91E63}  %
\definecolor{myBlue}{HTML}{2196F3}  %
\definecolor{myGreen}{HTML}{4CAF50} %

\colorlet{pinkBlueBlend}{myPink!50!myBlue}    %
\colorlet{pinkGreenBlend}{myPink!50!myGreen}  %
\colorlet{blueGreenBlend}{myBlue!50!myGreen}  %

\begin{table}[t]
    \centering
    \footnotesize
    \vspace{-1.5em}
    \begin{tabular}{l|>{\centering\arraybackslash}p{0.5cm} >{\centering\arraybackslash}p{0.65cm} |>{\centering\arraybackslash}p{0.65cm} >{\centering\arraybackslash}p{0.65cm} >{\centering\arraybackslash}p{0.65cm}}
         Model & CD$\downarrow$ & depth$\downarrow$ & cPSNR$\uparrow$& cSSIM$\uparrow$& cLPIPS$\downarrow$  \\
         \hline
         a. SV3D pretrained & 3.472 & 19.651 & 26.751 & 0.865 & 0.070 \\ %
         b. SV3D fine-tuned & 3.342 & 16.493 & 27.708 & 0.876 & 0.061 \\ %
         \rowcolor{myPink!20} %
         c. +RCN$^\dag$ & 3.212 & 15.671 & 28.257 & 0.889 & 0.055 \\ %
         \rowcolor{myPink!20}
         d. +CaP$^\dag$ & 3.244 & 15.528 & 27.980 & 0.884 & 0.057 \\ %
         \rowcolor{myPink!20}
         e. +RCN$^\dag$+Cap$^\dag$ & 3.246 & 15.369 & 27.809 & 0.884 & 0.058 \\ %
         \rowcolor{pinkBlueBlend!30}
         f. +RCN$^\dag$+Cov$^{\ddag}$ & {\bf 3.127} & {\bf 13.862} & 28.615 & 0.890 & 0.052 \\ %
         \rowcolor{myBlue!20}
         g. +CaP$^\dag$+DINOv2$^{\ddag}$ & 3.146 & 14.301 & 28.507 & 0.891 & 0.053 \\ %
         \rowcolor{blueGreenBlend!30}
         h. +CaP$^\dag$+Cov$^{\ddag}$ (\textbf{Ours}) & 3.154 & 14.204 & {\bf 28.934} & {\bf 0.897} & 0.052 \\ %
         \rowcolor{myGreen!22}
         i. +CaP$^\dag$+Cov$^{\ddag}$+sync$^\natural$ & 3.145 & 13.919 & 28.881 & 0.895 & {\bf 0.051} \\ %
    \end{tabular}
    \caption{\textbf{Investigating different design choices} added on top of SV3D~\cite{voletiSV3DNovelMultiview2024a}: camera embedding ($\dag$), input image encoder ($\ddag$), and multi-view feature syncronization ($\natural$). Results on GSO.}
    \label{tab:ablate-architecture}
\end{table}

\noindent\textbf{Input image encoder.} We use SV3D combined with CaP from EscherNet as our base model and replace the CLIP input image encoder with ConvNextV2 or DINOv2 encoder. We choose the encoder that produces exactly same feature dimension as the CLIP encoder used in SV3D. Hence the network architecture is exactly the same except for the feature used for cross attention. We fine tune the model for 50k steps to adapt to the new input image feature and results are reported in \cref{tab:ablate-architecture}. It can be seen that both DINOv2 (\cref{tab:ablate-architecture}g) and ConvNextV2 (\cref{tab:ablate-architecture}h) are better than the CLIP encoder (\cref{tab:ablate-architecture}b), while the difference between two vision encoders is small. We also show some results in \cref{fig:qualitative} where our model with ConvNextV2 encoder can preserve better details from input than CLIP based SV3D. Given that ConvNextV2 is more efficient than DINO model, we propose using ConvNextV2 to replace the CLIP encoder. 

\begin{table}[]
    \centering
    \vspace{-0.5em}
    \footnotesize
    \begin{tabular}{c| c|>{\centering\arraybackslash}p{0.65cm} >{\centering\arraybackslash}p{0.80cm} |>{\centering\arraybackslash}p{0.80cm} >{\centering\arraybackslash}p{0.8cm} >{\centering\arraybackslash}p{0.8cm}}
    \toprule[1.0pt]
         & \#Objs. & CD$\downarrow$ & depth$\downarrow$ & cPSNR$\uparrow$& cSSIM$\uparrow$& cLPIPS$\downarrow$ \\
         \hline
         \parbox[t]{1mm}{\multirow{3}{*}{\rotatebox[origin=c]{90}{GSO (syn.)}}} & 10k & 3.354 & 16.919 & 27.765 & 0.881 & 0.061 \\
        & 50k & 3.186 & 13.504 & 28.053 & 0.881 & 0.056 \\
        & 100k & \textbf{3.148} & \textbf{13.351} & 28.142 & 0.885 & 0.053 \\
        & 150k & 3.154 & 14.204 & \textbf{28.934} & \textbf{0.897} & \textbf{0.052} \\
         
         \midrule
         \parbox[t]{1mm}{\multirow{3}{*}{\rotatebox[origin=c]{90}{CO3D (real)}}} & 10k & 3.349 & 21.703 & 24.261 & 0.855 & 0.079 \\
& 50k & 3.176 & 17.556 & 25.118 & 0.866 & 0.068 \\
& 100k & 3.166 & 17.819 & 25.131 & 0.868 & 0.067 \\
& 150k & \textbf{3.099} & \textbf{16.942} & \textbf{25.994} & \textbf{0.880} & \textbf{0.062} \\
    \bottomrule[1.0pt]
    \end{tabular}
    \vspace{-0.3em}
    \caption{\textbf{The effect of training data amount} on 3D consistency when replacing CLIP with the convolution encoder. More data improves mainly the generalization to real images (CO3D~\cite{reizenstein2021co3d}). }
    \label{tab:ablate-data}
\end{table}

\noindent\textbf{Interaction between target views.}
SyncDreamer~\cite{liu2023syncdreamer} computes a synchronized 3D feature volume and derive multi-view images from it, leading to strong 3D consistency (\cref{tab:gso-vs-co3d}). 
We combine such 3D feature volume with the video based SV3D + CaP and ConvNextV2 encoding and results are shown in \cref{tab:ablate-architecture}. Surprisingly, the performance gain with the additional synchronized 3D convolution is small. We hypothesize that the explicit 3D feature interaction from SyncDreamer is already implicitly achieved by the dense spatial-temporal attention of SV3D. Hence \textbf{we adopt SV3D+Cap+Cov as the model for \texttt{Ours}}.

\noindent\textbf{Training data amount.} Another important aspect to consider is the amount of data needed. 
Since it is not computationally feasible to retrain every single method with different amounts of data, we study the effect of data amount on performance when adding a new module to a pretrained model. Specifically, we start with SV3D+CaP and replace the CLIP image encoder with a convolution image encoder and fine tune it on different amount of objaverse objects. The 3D consistency results are shown in \cref{tab:ablate-data}. It can be seen that 50K objects are sufficient to achieve good performance on synthetic images (GSO100) and improvement is small after that. 
By contrast, performance continues to improve with more data when testing on real images.

%% file: sec/6_conclusion.tex
\section{Conclusion}
We present \benchName{}, a comprehensive benchmark suite for evaluating multi-view generation models (MVGs). We unify 10 metric dimensions to evaluate the 3D consistency, image quality, and semantic consistency of generated images. Experiments show that our 3D consistency metric reports meaningful scores and fair comparisons of methods. Our quality and semantic consistency metrics align well with human perception: Pearson scores ranging from 0.69 to 0.92. We evaluate 12 SoTA MVGs and find that there is a trade-off between 3D consistency and image quality, and no method can achieve the best performance in all dimensions. We also observe that large performance gap persists between synthetic and real images, and most methods are not robust to different elevations, azimuths, or lightings. 

We investigate four key design choices of MVGs, including camera embedding, input image encoding, attention mechanism, and training data amount. We find that the convolution encoder can preserve fine details of the input image, resulting in better 3D consistency. Explicit 3D convolution does not provide much improvement on top of video-based models. While 50k training objects are sufficient for synthetic input, more data is necessary to improve generalization to real data. We will publicly release our benchmark suite and fine-tuned models.%

\noindent
{\footnotesize
\textbf{Acknowledgements.} We thank RVH group members \cite{rvh_grp} for their helpful discussions. This work is funded by the Deutsche Forschungsgemeinschaft (DFG, German Research Foundation) - 409792180 (Emmy Noether Programme,
project: Real Virtual Humans), and German Federal Ministry of Education and Research (BMBF): Tübingen AI Center, FKZ: 01IS18039A, and Amazon-MPI science hub. Gerard Pons-Moll is a Professor at the University of Tübingen endowed by the Carl Zeiss Foundation, at the Department of Computer Science and a member of the Machine Learning Cluster of Excellence, EXC number 2064/1 – Project number 390727645.
}

%% file: sec/X_suppl.tex
\clearpage
\setcounter{page}{1}
\maketitlesupplementary

In this supplementary, we first discuss the implementation details, including our \benchName{} metrics (\cref{supp:impl-metric}) and evaluation experiment setups (\cref{supp:impl-exps}). We then show additional evaluation result and analysis in \cref{supp:additional-results}, and conclude with discussion of limitations.

\section{Implementation Details}
We discuss the details of our metrics and experiment setups. Our benchmark suite and pre-trained models will be publicly released. 
\subsection{\benchName{} Metric Implementation}\label{supp:impl-metric}
\paragraph{View sets split.} For 3D consistency metric, we split the generated multi-view images into two subsets and fit 3DGS separately to them. We allow small overlap when the total number of generated views is small. There are three different number of output views for all the methods we evaluated: 16, 18, 21, see \cref{tab:best-setup}. The view indices for the two subsets are: 1). Output 16 views: [0, 2, 4, 5, 6, 8, 9, 10, 11, 12, 14], [1, 3, 5, 6, 7, 9, 11, 12, 13, 14, 15]. 2). Output 18 views: [0, 1, 2, 4, 6, 8, 10, 12, 14, 16], [0, 1, 2, 3, 5, 7, 9, 11, 13, 15, 17]. 3). Output 21 views: the first (input) view is excluded and rest is divided into two non-overlap views, namely [0,  2,  4,  6,  8, 10, 12, 14, 16, 18], [ 1,  3,  5,  7,  9, 11, 13, 15, 17, 19]. 

\paragraph{3DGS optimization.} We use the original version of 3DGS~\cite{kerbl3Dgaussians} for optimization. We explored more advanced version of 3DGS but found that they are either less accurate for object level multi-views \cite{Huang2DGS2024, Yu2024MipSplatting, sigg24taming3dgs} or the runtime is too long \cite{Yu2024GOF}. We hence stick to the original 3DGS version and randomly sample 100k points from unit cube [-1, 1] to initialize the Gaussians and optimize for 10k steps. White background is used during optimization as all methods generate images with white background. 

\paragraph{Test view rendering.} We render the optimized 3DGSs into RGB and depth images to compute the depth, cPSNR, cSSIM, cLPIPS metrics. To produce comparable numbers, the test views have to be the same for two 3DGSs and across all methods, for the same test object. The test views should be diverse so that is does not favor output elevation angles specific to some methods while it should also be close to the views used to fit 3DGS, otherwise the calculated scores are dominated by 3DGS fitting error instead of multi-view inconsistency. To this end, we use two setups to choose the views for rendering: a). Random views sampled from a fixed range, and b). Fixed views that differ 15 elevation degrees from generated multi-views. For both setup $K=16$ views are used for rendering, and each object might have different test views but the same views are always used across methods for fair comparison. 

\noindent\textbf{Random test views} are used for best setup performance, robustness w.r.t to lighting and azimuth conditions. As existing methods generate multi-view images with elevation angle ranging from 0 to 30 degrees (\cref{tab:best-setup}), we uniformly sample elevation from range [-15, 45], azimuth from [0, 360], and camera distance from [1.5, 1.9]. The field of view (Fov) is fixed at 42 degree such that it does not favor any of the methods evaluated. 

\noindent\textbf{Fixed test views} are used for generalization to real images and robustness w.r.t to different elevation degrees. In these setups, the output elevation differ a lot and it is difficult to define a common range where 3DGS fitting also works well and we can sample elevation from. To this end, we take 8 views with equal azimuth distance from 8.5 to 360 degree and the elevation is 15 degree higher than the elevation of generated multi-views. The other 8 views have the same azimuth angles but the elevation is 15 degree lower than the output multi-view elevation. The fov and camera distance are fixed to 42 degree and 3.2m. We choose these azimuth, fov, and distance to not favor any methods. Note that this will lead to consistency scores that are not comparable across different output elevations, which address next.

\paragraph{Normalization of the consistency scores.} The exact scores of our consistency metrics depend on the views used to render test images and the raw numbers are not directly comparable if the views are different. This is the case when we want to evaluate the robustness of a method w.r.t to different elevation angles (see discussion above). We hence propose to normalize the raw numbers using the upper bound scores obtained from ground truth multi-view images. Let $e_\text{gt}, e_\text{mvg}$ be the raw consistency score defined in \cref{subsec:3d-consistency-metric} using MVG and GT images of the same camera views. The normalized error $e_\text{mvg, n}$ is computed as:
\begin{equation}
    e^i_\text{mvg, n} =
\begin{cases} 
    \frac{e_\text{mvg}}{e_\text{gt}} & \text{if type(} e \text{)} \in \text{ \{cPSNR, cSSIM\}}, \\
    1 - \frac{e_\text{mvg}}{\max e_\text{mvg}} &  \text{if type(} e \text{)} \in \text{ \{CD, depth, cLPIPS\}}.
\end{cases}
\label{eq:metric-normalization}
\end{equation}
here $\max e_\text{mvg}$ is the maximum error for this metric among our evaluated methods. This yields a score between 0 and 1 and it is always the higher the better. This normalization is also used to visualize the bottom plots in \cref{fig:teaser}.

\paragraph{Prompt templates for VLM based metrics.} We propose four metrics based on the pretrained 73B InternVL2.5 VLM~\cite{chen2024internvl_2_5}. The same model is used to obtain the reference attributes (\cref{subsec:semantic-quality}) via three-round prompts given multi-view images. The three sequential prompts are: 1). ``Here are images of a daily object, what is the appearance style of this object? Ignore the background, focus on the appearance, style and design instead of describing the object type, return the appearance style only and in less than 5 words.", 2). ``Which object it is? Just return the class name, do not repeat question. Use daily used common words. If there are multiple possibilities, return like this: classname 1 or classname2 or classname3...", 3). ``What is the main color(s) of this object? simply answer the color(s), summarize to less than 4 colors."

We then use the answers from these prompts as the reference attributes and evaluate the semantic consistency using the following templates: 1). \texttt{class}: ``Is $[\text{obj cls}]$ presented in this image? just answer yes or no." 2). \texttt{color}: ``Does the object (possibly $[\text{obj cls}]$) shown in this image have the color(s): $[\text{color}]$? just answer yes or no. 3). ``Is the appearance style of the object (possibly $[\text{obj cls}]$): $[\text{style}]$? just answer yes or no." 

We also use the same model to asses the image quality (IQ-vlm) which we find align well with human perception. The prompt template is: ``Is this image an overall high-quality image with good overall structure, good visual quality, nice color harmony, clear object and free of strange artifacts and distortions? just answer yes or no.".

\paragraph{Runtime performance.} The most compute expensive steps in our evaluation pipeline are 3DGS optimization and VLM assessment, which takes around 76s (two subsets) and 12s per input image to finish on L40s GPU. In total it takes around 2.7 hours to evaluate 100 objects which is still reasonable. More advanced techniques such as better 3DGS initialization~\cite{chen2024v3d} or VLM inference via API call could be adopted to speed up evaluation. We leave these for future works.

\subsection{Experiment Setups}\label{supp:impl-exps}
\begin{figure}[t]
    \centering
    \includegraphics[width=1.0\linewidth]{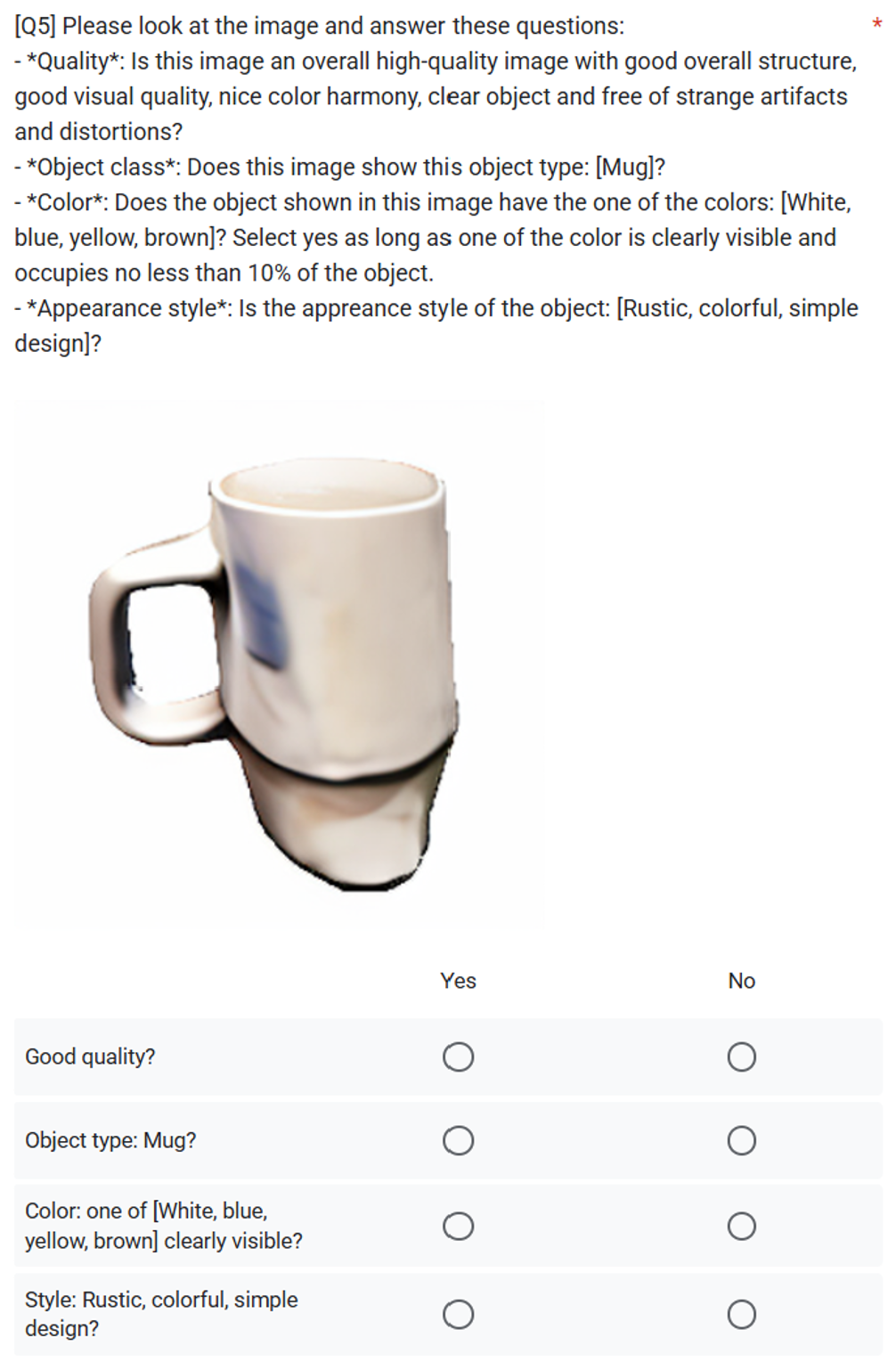}
    \caption{Example question from our user study on the alignment between our VLM based metrics and human preference.}
    \label{fig:user-study-vlm}
\end{figure}

\begin{figure}[t]
    \centering
    \includegraphics[width=1.0\linewidth]{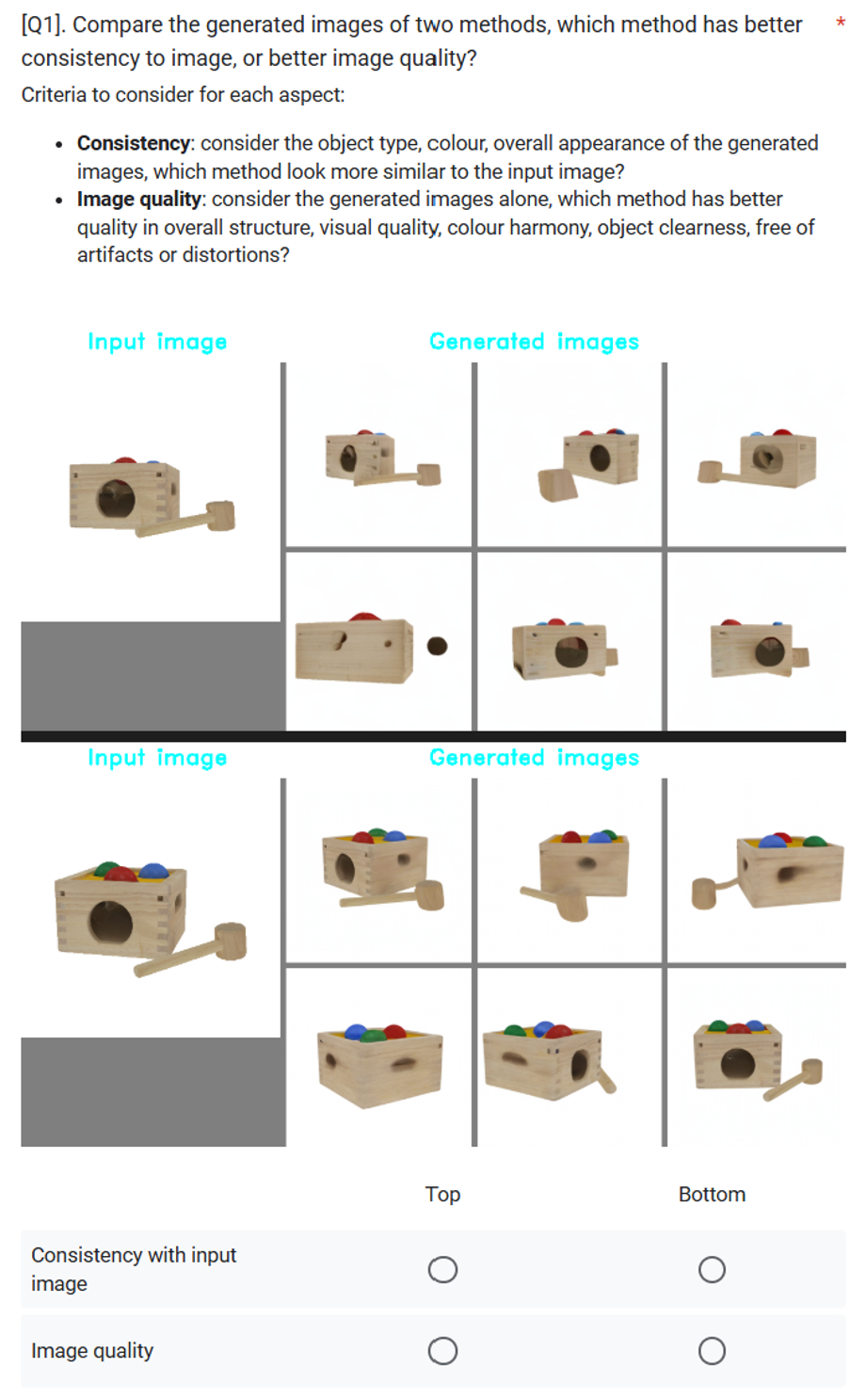}
    \caption{Example question from our user study on the alignment between our oFID score and human preference.}
    \label{fig:user-study-oFID}
\end{figure}

\paragraph{User study.} We conduct user studies to verify our oFID score and VLM based metrics, each with 400 questions answered by 10 users. As 400 questions are too many for one single user study survey, we divide it into 8 smaller surveys, each with 50 questions. We then recruit 10 users to finish each survey and no overlap is allowed for different surveys. Hence in total we have 80 different users to participate one user study. This ensures sufficient diversity and statistically meaningful results. We show example questions from our user studies in \cref{fig:user-study-vlm} and \cref{fig:user-study-oFID}. 

\paragraph{Evaluation setup for existing methods.} We show the input and output setups for all the methods we evaluate in the \textbf{best-setup performance} experiment in \cref{tab:best-setup}. We use ambience light of 1.0 and zero azimuth angle for this setup. The rendering setup is the same for \textbf{robustness evaluation} except for the attribute we want to evaluate (elevation, light intensity, and azimuth angle). For \textbf{generalization to real images}, we cannot control the rendering anymore hence we use the same input image crop for different methods, which has 0.2 margin from the object bounding box to image boarder. The number of output views of each method remain the same as in best-setup evaluation.  

\paragraph{MVG design choice experiments.} We use the 150k kiui objects filtered by LGM~\cite{tang2024lgm} as our training data. Following same camera parameters in SV3D~\cite{voletiSV3DNovelMultiview2024a}, we render each object from 84 views and randomly pick 21 views at training time. We adopt the dynamic orbit rendering from SV3D which adds perturbations of azimuth and elevation angles to the equally distributed static orbit. We pre-compute the latent features of CLIP~\cite{radford2021clip}, SVD~\cite{rombach2021latentdiffusion}, DINOV2~\cite{oquab2024dinov2} and ConvNextV2~\cite{Woo2023ConvNeXtV2} to speed up training. We use batch size of 64, learning rate of 2e-5 for all the experiments. The total training steps is 26k for camera embedding experiments and 50k for all other experiments.

\section{Additional Results and Analysis}\label{supp:additional-results}
\paragraph{Full evaluation results.} We show all scores of our \benchName{} from all evaluated methods on four datasets in \cref{tab:eval-gso100v2} (GSO~\cite{downs2022gso}), \cref{tab:eval-omni202} (Omni3D~\cite{wu2023omniobject3d}), \cref{tab:eval-co3d} (CO3D~\cite{reizenstein2021co3d}), and \cref{tab:eval-mvimgnet} (MVImgnet~\cite{yu2023mvimgnet}). It can be seen that our method achieves the best overall 3D consistency and on par performance on image quality and semantic consistency.

\begin{figure}
    \centering
    \includegraphics[width=1.0\linewidth]{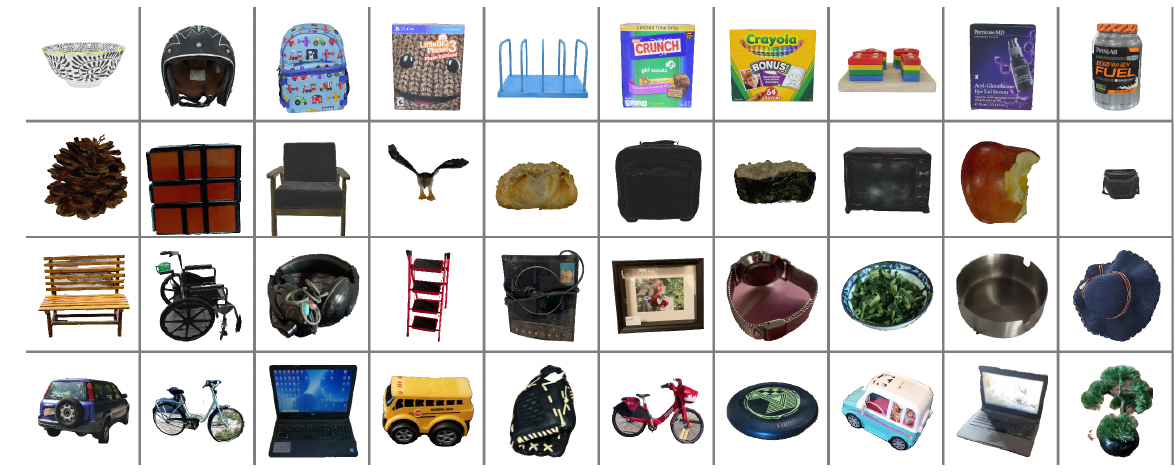}
    \caption{\textbf{The most challenging test images} from GSO\cite{downs2022gso}, Omni3D\cite{wu2023omniobject3d}, MVImgnet\cite{yu2023mvimgnet} and CO3D\cite{reizenstein2021co3d}. Methods produce the most 3D inconsistent images for these inputs due to their complex geometric structure or high frequency details.}
    \label{fig:worst10-examples}
\end{figure}

\begin{figure}
    \centering
    \includegraphics[width=1.0\linewidth]{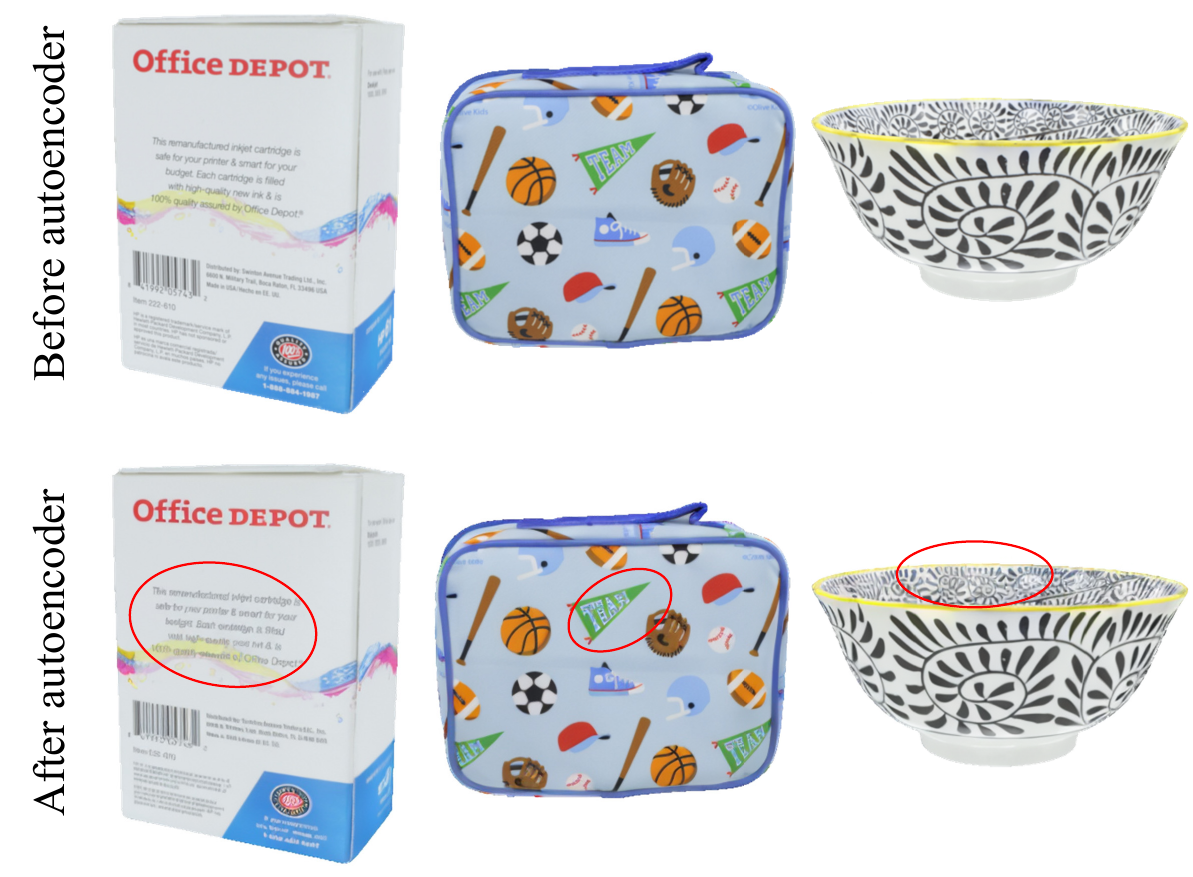}
    \caption{Degradation of image quality after passing through the autoencoder of SV3D~\cite{voletiSV3DNovelMultiview2024a}. Clearly the high frequency details are destroyed by the autoencoder.}
    \label{fig:supp-autoencoder}
\end{figure}

\paragraph{Methods struggle with fine-grained details.} We rank the input images based on the 3D consistency score (cPSNR) from different methods and visualize the 10 common inputs that have the worst scores in \cref{fig:worst10-examples}. It can be seen that the common challenging images are the objects with complex and fine-grained geometric structures and textures such as bicycle, flowers, text boxes. Diving further into this problem, we find that the autoencoder used in all MVGs already destroys the high frequency structures after one single pass through the autoencoder. We show two examples in \cref{fig:supp-autoencoder}. To further understand the effect on 3D consistency, we send the ground truth multi-view images of 30 GSO objects~\cite{kongEscherNetGenerativeModel} through the autoencoder of SV3D~\cite{voletiSV3DNovelMultiview2024a}. The consistency scores before and after the autoencoder are (CD / depth / cPSNR / cSSIM / cLPIPS): 2.58 / 9.82 / 31.94 / 0.94 / 0.03 (before), 2.69 / 9.05 / 30.29 / 0.92 / 0.04 (after). It can be clearly seen that the 3D texture consistency scores already degrade after single feedforward through the autoencoder. Interestingly, the depth error, sensitive to inconsistency in edges, decreases which indicates the images are indeed smoother with high-frequency details lost.

\begin{table}[]
    \centering
    \begin{tabular}{p{2.8cm}|>{\centering\arraybackslash}p{0.55cm} >{\centering\arraybackslash}p{0.55cm}>{\centering\arraybackslash}p{0.70cm}>{\centering\arraybackslash}p{1.80cm}}
         Methods & Fov & Elev. & Dist.(m) & \#Out views \\
         \hline
         SyncDreamer based & \multirow{2}{*}{49.1} & \multirow{2}{*}{30} & \multirow{2}{*}{1.5} & \multirow{2}{*}{16}\\
         \cite{huMVDFusionSingleview3D, liu2023syncdreamer, zhengFree3DConsistentNovel2024, huangEpiDiffEnhancingMultiView, yang2024hi3d} & & & &   \\ 
         V3D \cite{chen2024v3d} & 60 & 0 & 2.0 & 18  \\ 
         SV3D \cite{voletiSV3DNovelMultiview2024a}, ours & 33.8 & 12.5 & 2.35 & 21\\ 
         Zero123 based & \multirow{2}{*}{49.1} & \multirow{2}{*}{0} & \multirow{2}{*}{1.85} & \multirow{2}{*}{21} \\ 
         \cite{stablezero123, deitke2023objaversexl, kongEscherNetGenerativeModel, yangViewFusionMultiViewConsistency, kwakViVid1to3NovelView} & & & & \\
         
    \end{tabular}
    \caption{The input rendering (Fov, camera elevation degree and distance) and number of output views of each method for the best performance evaluation. }
    \label{tab:best-setup}
\end{table}

\begin{table*}[ht]
    \centering
    \small
    \begin{tabular}{l|c c | c c c| c c | c c c }
        \multirow{2}{*}{Method} & \multicolumn{2}{c}{Geometry consistency} & \multicolumn{3}{c}{Texture consistency} & \multicolumn{2}{c}{Image quality} & \multicolumn{3}{c}{Semantic consistency} \\
        & CD $\downarrow$ & depth $\downarrow$ & PSNR $\uparrow$& SSIM $\uparrow$& LPIPS $\downarrow$ & FID $\downarrow$ & \qualityMetric{} $\uparrow$& class $\uparrow$& color $\uparrow$& style $\uparrow$ \\
        \hline

Ours & 3.15 & \textbf{14.20} & \textbf{28.93} & \textbf{0.90} & \textbf{0.05} & 20.46 & 0.82 & 0.86 & 0.94 & 0.93 \\
SyncDreamer\cite{liu2023syncdreamer} & \textbf{2.99} & 17.29 & 26.83 & 0.87 & 0.07 & 22.72 & 0.53 & 0.84 & 0.96 & 0.94 \\
SV3D-tune & 3.34 & 16.49 & 27.71 & 0.88 & 0.06 & 19.06 & 0.80 & 0.89 & 0.95 & \textbf{0.96} \\
SV3D\cite{voletiSV3DNovelMultiview2024a} & 3.47 & 19.65 & 26.75 & 0.86 & 0.07 & 21.31 & 0.77 & 0.85 & 0.92 & 0.93 \\
Hi3D\cite{yang2024hi3d} & 3.29 & 21.69 & 24.60 & 0.84 & 0.09 & 18.68 & \textbf{0.87} & \textbf{0.89} & 0.95 & 0.95 \\
V3D\cite{chen2024v3d} & 4.25 & 28.08 & 23.84 & 0.81 & 0.12 & 21.20 & 0.77 & 0.86 & 0.96 & 0.91 \\
EscherNet\cite{kongEscherNetGenerativeModel} & 4.34 & 20.61 & 23.89 & 0.79 & 0.11 & 24.71 & 0.57 & 0.77 & 0.90 & 0.88 \\
MVDFusion\cite{huMVDFusionSingleview3D} & 4.77 & 38.74 & 21.44 & 0.76 & 0.15 & 25.60 & 0.48 & 0.88 & 0.94 & 0.94 \\
ViewFusion\cite{yangViewFusionMultiViewConsistency} & 5.33 & 40.20 & 22.34 & 0.80 & 0.14 & 22.03 & 0.63 & 0.82 & 0.92 & 0.92 \\
EpiDiff\cite{huangEpiDiffEnhancingMultiView} & 5.77 & 50.65 & 20.28 & 0.72 & 0.19 & \textbf{16.53} & 0.77 & 0.89 & \textbf{0.97} & 0.94 \\
Free3D\cite{zhengFree3DConsistentNovel2024} & 6.03 & 44.27 & 20.26 & 0.77 & 0.18 & 27.30 & 0.73 & 0.78 & 0.82 & 0.90 \\
Vivid123\cite{kwakViVid1to3NovelView} & 7.57 & 43.97 & 21.74 & 0.81 & 0.18 & 38.91 & 0.63 & 0.66 & 0.78 & 0.80 \\
Zero123\cite{stablezero123} & 10.99 & 63.72 & 17.37 & 0.67 & 0.29 & 21.35 & 0.73 & 0.82 & 0.90 & 0.93 \\
Zero123-xl\cite{deitke2023objaversexl} & 15.40 & 68.13 & 17.10 & 0.66 & 0.30 & 20.72 & 0.72 & 0.83 & 0.91 & 0.94 \\

    \end{tabular}
    \caption{Best setup performance on the GSO~\cite{downs2022gso} dataset.}
    \label{tab:eval-gso100v2}
\end{table*}

\begin{table*}[ht]
    \centering
    \small
    \begin{tabular}{l|c c | c c c| c c | c c c }
        \multirow{2}{*}{Method} & \multicolumn{2}{c}{Geometry consistency} & \multicolumn{3}{c}{Texture consistency} & \multicolumn{2}{c}{Image quality} & \multicolumn{3}{c}{Semantic consistency} \\
        & CD $\downarrow$ & depth $\downarrow$ & PSNR $\uparrow$& SSIM $\uparrow$& LPIPS $\downarrow$ & oFID $\downarrow$ & \qualityMetric{} $\uparrow$& class $\uparrow$& color $\uparrow$& style $\uparrow$ \\
        \hline
Ours & 2.98 & \textbf{11.63} & \textbf{29.09} & \textbf{0.92} & \textbf{0.04} & 15.47 & \textbf{0.54} & 0.77 & 0.85 & \textbf{0.88} \\
SyncDreamer\cite{liu2023syncdreamer} & \textbf{2.93} & 13.60 & 27.24 & 0.89 & 0.06 & 5.94 & 0.24 & 0.64 & 0.85 & 0.79 \\
Hi3D\cite{yang2024hi3d} & 3.13 & 17.63 & 25.25 & 0.88 & 0.08 & 16.07 & 0.55 & 0.74 & 0.87 & 0.84 \\
SV3D-tune & 3.16 & 14.11 & 27.69 & 0.91 & 0.05 & 15.00 & 0.51 & 0.79 & 0.88 & 0.89 \\
SV3D\cite{voletiSV3DNovelMultiview2024a} & 3.46 & 19.46 & 26.02 & 0.88 & 0.07 & 17.60 & 0.50 & 0.69 & 0.85 & 0.85 \\
V3D~\cite{chen2024v3d} & 4.51 & 23.62 & 23.01 & 0.85 & 0.12 & 17.70 & 0.46 & 0.70 & 0.84 & 0.85 \\
EscherNet\cite{kongEscherNetGenerativeModel} & 5.01 & 23.25 & 21.87 & 0.77 & 0.14 & 22.39 & 0.41 & 0.60 & 0.80 & 0.79 \\
MVDFusion\cite{huMVDFusionSingleview3D} & 5.67 & 47.96 & 19.04 & 0.76 & 0.19 & 26.89 & 0.21 & 0.69 & 0.83 & 0.82 \\
EpiDiff\cite{huangEpiDiffEnhancingMultiView} & 6.78 & 57.31 & 18.37 & 0.73 & 0.21 & \textbf{14.61} & 0.52 & \textbf{0.79} & \textbf{ 0.89} & 0.88 \\
ViewFusion\cite{yangViewFusionMultiViewConsistency} & 7.88 & 54.32 & 17.90 & 0.73 & 0.24 & 16.96 & 0.44 & 0.68 & 0.85 & 0.87 \\
Free3D\cite{zhengFree3DConsistentNovel2024} & 8.02 & 52.58 & 16.97 & 0.72 & 0.25 & 23.78 & 0.50 & 0.61 & 0.77 & 0.79 \\
Zero123-xl\cite{deitke2023objaversexl} & 13.67 & 70.17 & 13.64 & 0.60 & 0.39 & 17.86 & 0.51 & 0.67 & 0.84 & 0.86 \\
Zero123\cite{stablezero123} & 14.17 & 70.32 & 14.15 & 0.62 & 0.38 & 17.62 & 0.51 & 0.69 & 0.84 & 0.88 \\
Vivid123\cite{kwakViVid1to3NovelView} & 14.31 & 56.07 & 17.80 & 0.76 & 0.26 & 27.98 & 0.50 & 0.56 & 0.74 & 0.74 \\
    \end{tabular}
    \caption{Best setup performance on the Omni3D~\cite{wu2023omniobject3d} dataset.}
    \label{tab:eval-omni202}
\end{table*}

\begin{table*}[t!]
    \centering
    \small
    \begin{tabular}{l|c c | c c c| c c | c c c }
        \multirow{2}{*}{Method} & \multicolumn{2}{c}{Geometry consistency} & \multicolumn{3}{c}{Texture consistency} & \multicolumn{2}{c}{Image quality} & \multicolumn{3}{c}{Semantic consistency} \\
        & CD $\downarrow$ & depth $\downarrow$ & PSNR $\uparrow$& SSIM $\uparrow$& LPIPS $\downarrow$ & FID $\downarrow$ & \qualityMetric{} $\uparrow$& class $\uparrow$& color $\uparrow$& style $\uparrow$ \\
        \hline

Ours & 3.10 & 16.94 & \textbf{25.99} & \textbf{0.88} & \textbf{0.06} & 23.40 & 0.29 & 0.80 & \textbf{0.86} & 0.82 \\
SyncDreamer\cite{liu2023syncdreamer} & \textbf{3.04} & \textbf{13.48} & 25.30 & 0.88 & 0.06 & 30.96 & 0.12 & 0.69 & 0.83 & 0.70 \\
SV3D-tune & 3.43 & 19.99 & 24.32 & 0.85 & 0.08 & 21.71 & 0.26 & 0.82 & 0.84 & 0.83 \\
SV3D\cite{voletiSV3DNovelMultiview2024a} & 3.48 & 25.80 & 23.72 & 0.87 & 0.13 & 24.19 & 0.29 & 0.76 & 0.87 & 0.78 \\
EscherNet\cite{kongEscherNetGenerativeModel} & 5.14 & 26.46 & 20.34 & 0.71 & 0.14 & 28.54 & 0.26 & 0.71 & 0.79 & 0.72 \\
Hi3D\cite{yang2024hi3d} & 5.60 & 31.09 & 20.92 & 0.81 & 0.12 & 25.51 & 0.35 & 0.75 & 0.82 & 0.75 \\
MVDFusion\cite{huMVDFusionSingleview3D} & 5.77 & 47.43 & 17.50 & 0.71 & 0.20 & 27.16 & 0.19 & 0.75 & 0.82 & 0.78 \\
EpiDiff\cite{huangEpiDiffEnhancingMultiView} & 7.71 & 58.58 & 15.66 & 0.64 & 0.26 & \textbf{20.58} & 0.31 & 0.84 & 0.86 & 0.82 \\
ViewFusion\cite{yangViewFusionMultiViewConsistency} & 7.75 & 49.76 & 16.49 & 0.77 & 0.29 & 22.10 & 0.33 & 0.82 & 0.85 & 0.82 \\
Vivid123\cite{kwakViVid1to3NovelView} & 9.81 & 56.38 & 15.31 & 0.69 & 0.29 & 35.89 & \textbf{0.49} & 0.70 & 0.76 & 0.72 \\
V3D\cite{chen2024v3d} & 10.45 & 58.71 & 16.39 & 0.71 & 0.26 & 28.76 & 0.32 & 0.72 & 0.85 & 0.77 \\
Free3D\cite{zhengFree3DConsistentNovel2024} & 11.15 & 60.95 & 14.42 & 0.76 & 0.33 & 32.84 & 0.32 & 0.71 & 0.75 & 0.75 \\
Zero123\cite{stablezero123} & 12.06 & 64.74 & 13.16 & 0.55 & 0.38 & 21.22 & 0.38 & 0.84 & 0.87 & \textbf{0.86} \\
Zero123-xl\cite{deitke2023objaversexl} & 12.58 & 66.99 & 12.97 & 0.54 & 0.38 & 20.83 & 0.34 & \textbf{0.85} & 0.86 & 0.84 \\

    \end{tabular}
    \caption{Evaluation results on the CO3D\cite{reizenstein2021co3d} dataset with manually selected front view and annotated elevation angles.}
    \label{tab:eval-co3d}
\end{table*}

\begin{table*}[t]
    \centering
    \small
    \begin{tabular}{l|c c | c c c| c c | c c c }
        \multirow{2}{*}{Method} & \multicolumn{2}{c}{Geometry consistency} & \multicolumn{3}{c}{Texture consistency} & \multicolumn{2}{c}{Image quality} & \multicolumn{3}{c}{Semantic consistency} \\
        & CD $\downarrow$ & depth $\downarrow$ & PSNR $\uparrow$& SSIM $\uparrow$& LPIPS $\downarrow$ & FID $\downarrow$ & \qualityMetric{} $\uparrow$& class $\uparrow$& color $\uparrow$& style $\uparrow$ \\
        \hline

Ours & 3.04 & 17.58 & \textbf{26.43} & \textbf{0.88} & \textbf{0.06} & 22.10 & \textbf{0.37} & \textbf{0.74} & \textbf{0.88} & 0.84 \\
SyncDreamer\cite{liu2023syncdreamer} &\textbf{ 2.87} & \textbf{15.35} & 25.44 & 0.88 & 0.06 & 30.64 & 0.17 & 0.59 & 0.84 & 0.79 \\
SV3D-tune & 3.37 & 21.94 & 24.97 & 0.85 & 0.08 & 22.04 & 0.34 & 0.72 & 0.88 & 0.82 \\
SV3D\cite{voletiSV3DNovelMultiview2024a} & 3.39 & 26.17 & 23.99 & 0.83 & 0.09 & 22.07 & 0.33 & 0.71 & 0.87 & 0.79 \\
EscherNet\cite{kongEscherNetGenerativeModel} & 5.35 & 29.55 & 20.31 & 0.72 & 0.15 & 25.78 & 0.32 & 0.66 & 0.82 & 0.78 \\
Hi3D\cite{yang2024hi3d} & 6.24 & 33.72 & 21.42 & 0.81 & 0.12 & 25.77 & 0.41 & 0.63 & 0.83 & 0.78 \\
MVDFusion\cite{huMVDFusionSingleview3D} & 6.29 & 50.54 & 17.27 & 0.70 & 0.22 & 28.45 & 0.26 & 0.62 & 0.80 & 0.78 \\
EpiDiff\cite{huangEpiDiffEnhancingMultiView} & 8.05 & 61.91 & 15.85 & 0.64 & 0.27 & \textbf{20.21} & 0.42 & 0.74 & 0.86 & 0.84 \\
ViewFusion\cite{yangViewFusionMultiViewConsistency} & 8.26 & 54.27 & 16.14 & 0.63 & 0.28 & 21.77 & 0.39 & 0.72 & 0.84 & 0.84 \\
Free3D\cite{zhengFree3DConsistentNovel2024} & 10.64 & 60.00 & 14.77 & 0.65 & 0.33 & 33.58 & 0.38 & 0.60 & 0.70 & 0.75 \\
Vivid123\cite{kwakViVid1to3NovelView} & 10.67 & 58.06 & 15.81 & 0.69 & 0.30 & 35.84 & \textbf{0.56} & 0.56 & 0.68 & 0.75 \\
V3D~\cite{chen2024v3d} & 10.74 & 65.40 & 16.30 & 0.71 & 0.26 & 27.89 & 0.40 & 0.65 & 0.79 & 0.79 \\
Zero123-xl\cite{deitke2023objaversexl} & 12.04 & 67.08 & 13.45 & 0.55 & 0.38 & 20.51 & 0.41 & 0.74 & 0.85 & \textbf{0.87} \\
Zero123\cite{stablezero123} & 12.11 & 66.76 & 13.61 & 0.56 & 0.38 & 20.83 & 0.42 & 0.74 & 0.85 & 0.86 \\

    \end{tabular}
    \caption{Evaluation results on the MVImgNet~\cite{yu2023mvimgnet} dataset with manually selected front view and annotated elevation angles.}
    \label{tab:eval-mvimgnet}
\end{table*}

\section{Limitations and Future Work}
We present the first comprehensive benchmark to evaluate 3D consistency of object multi-view generation models. Despite robust to various settings, there are still limitations of our benchmark. First, our method cannot evaluate methods that generate very few views ($<$10) as the 3DGS fitting is very inaccurate and fitting error instead of multi-view inconsistency dominates our consistency scores. One possible solution is to replace 3DGS fitting with pre-trained models that can take few-views as input and directly regress 3DGS, such as LGM~\cite{tang2024lgm}. This however requires the model to be robust to diverse camera setups which is still an ongoing research. Second, we curated four datasets which covers mainly daily objects, and most of them are indoor. It would be interesting to also consider outdoor objects such as buildings, statues or complex compositional shapes such as human-human or human-object interactions. Last but not least, we evaluate the robustness w.r.t lighting, elevation and azimuth angles. Real life objects have much more attributes that can affect the performance, such as the material, shading condition, specific object categories. One can do more comprehensive analysis could be done using our proposed metrics to understand the progress of SoTA methods. We leave these for future works.